\documentclass[11pt]{article}
\pdfoutput=1

\newlength{\abstractwidth}
\usepackage{bm}
\usepackage{changepage}
\usepackage{cite}
\usepackage{showlabels}
\usepackage{enumitem}
\usepackage{ifpdf}
 \usepackage{lmodern}
\usepackage[T1]{fontenc}
\usepackage[ansinew]{inputenc}
\usepackage{amsthm}
\usepackage{amsmath}
\usepackage{relsize}
\usepackage{amssymb}
\usepackage{comment}
\usepackage{tikz}
\usetikzlibrary{tikzmark}
\usepackage{graphicx}
\usepackage{color}
\definecolor{darkblue}{cmyk}{0.9,0.9,0,0}
\definecolor{darkgreen}{rgb}{0,0.55,0}
\definecolor{vert}{rgb}{0.1367,0.543,0.1367}
\usepackage[colorlinks=true,linkcolor=darkblue,citecolor=darkblue,urlcolor=darkblue]{hyperref}
\usepackage{epsfig}
\usepackage{epstopdf}
\usepackage{mathtools}
\usepackage{graphicx}

\usetikzlibrary{decorations.pathmorphing}

\usepackage{empheq} 
\usepackage{url}

\makeatletter
\long\def\@makecaption#1#2{
  \vskip\abovecaptionskip
  \sbox\@tempboxa{{\captionfonts #1: #2}}
  \ifdim \wd\@tempboxa >\hsize
    {\captionfonts #1: #2\par}
  \else
    \hbox to\hsize{\hfil\box\@tempboxa\hfil}
  \fi
  \vskip\belowcaptionskip}
\makeatother

\usepackage{enumitem}

\def\sl{SL(2,\Z)}

\def\L{\Lambda}

\def\z{\zeta}

\renewcommand{\thanks}[1]{\footnote{#1}}
\newcommand{\starttext}{
\setcounter{footnote}{0}
\renewcommand{\thefootnote}{\arabic{footnote}}}

\newcommand{\bea}{\begin{eqnarray}}
\newcommand{\eea}{\end{eqnarray}}
\newcommand{\be}{\begin{eqnarray}}
\newcommand{\ee}{\end{eqnarray}}
\newcommand{\bma}{\begin{matrix}}
\newcommand{\ema}{\cr\end{matrix}}
\newcommand{\<}{\langle}
\renewcommand{\>}{\rangle}

\def\cF{{\cal F}}
\def\cG{{\cal G}}

\def\cN{{\cal N}}
\def\cO{{\cal O}}

\def\cQ{{\cal Q}}

\def\Re{{\rm Re \,}}
\def\Im{{\rm Im \,}}

\def\Tr{{\rm Tr}}

\def\half{{1\over 2}}
\def\p{\partial}

\def\a{\alpha}

\def\g{\gamma}
\def\G{\Gamma}
\def\l{\lambda}
\def\eps{\epsilon}

\def\Li{{\rm Li}}

\def\no{\nonumber}

\def\({\left(}
\def\){\right)}
\def\[{\left[}
\def\]{\right]}

\def\<{\langle}
\def\>{\rangle}

\def\Z{\mathbb{Z}}
\def\R{\mathbb{R}}

\def\x{\times}

\DeclareMathOperator*{\Res}{Res}

\def\eps{\epsilon}

\def\1{{\rm 1-loop}}

\def\Tr{{\rm Tr}}

\def\c{\cite}

\def\zb{\bar z}
\def\cG{\mathcal{G}}
\def\cN{\mathcal{N}}

\def\G{\Gamma}
\def\p{\partial}
\def\o{\over}
\def\g{\gamma}
\def\D{\Delta}
\def\rar{\rightarrow}

\def\eqr{\eqref}

\def\O{{\cal O}}

\def\ssec{\subsection}
\def\sssec{\subsubsection}
\def\sec{\section}
\def\i{\infty}
\def\foot{\footnote}
\newcommand{\es}[2] {\begin{equation} \label{#1} \begin{split} #2 \end{split} \end{equation}}
\newcommand{\e}[2] {\begin{equation} \label{#1} #2 \end{equation}}

\def\t{\tau}

\usepackage{varioref}
\usepackage{makeidx}
\makeindex

\usepackage[english]{babel}
\usepackage{caption}
\usepackage{subfig}

\usepackage{tabularx}
 \renewcommand{\baselinestretch}{1.03}
 
\begin{document}
\starttext
\thispagestyle{empty}

\begin{flushright}
\end{flushright}

\vskip 1in

\begin{center}

{\LARGE \bf Exact Large Charge in $\cN=4$ SYM\\\vskip .05 in and Semiclassical String Theory}

\vskip 0.5in

{Hynek Paul, Eric Perlmutter, Himanshu Raj } 
   
\vskip 0.15in

{\small Universit\'e Paris-Saclay, CNRS, CEA, Institut de Physique Th\'eorique, 91191, Gif-sur-Yvette, France}

\vskip 0.15in

{\tt \small hynek.paul,perl,himanshu.raj@ipht.fr}

\vskip 0.8in

\begin{abstract}

\vskip 0.1in

In four-dimensional $\mathcal{N}=4$ super Yang-Mills theory with gauge group $SU(N)$, we present a closed-form solution for a family of integrated four-point functions involving stress tensor multiplet composites of arbitrary R-charge. These integrated correlators are shown to be equivalent to a one-dimensional semi-infinite lattice of harmonic oscillators with nearest-neighbor interactions, evolving over the fundamental domain of $SL(2,\mathbb{Z})$. The solution, exceptionally simple in an $SL(2,\mathbb{Z})$-invariant eigenbasis, is exact in the R-charge $p$, rank $N$, and complexified gauge coupling $\tau$. This permits a systematic and non-perturbative large charge expansion for any $N$ and $\tau$. Especially novel is a double-scaled ``gravity regime'' in which $p \sim N^2 \gg 1$, holographically dual to a large charge regime of semiclassical type IIB string theory in AdS$_5\, \times$ S$^5$. Our results in this limit provide a holographic computation of integrated semiclassical string amplitudes at arbitrary string coupling, including an emergent string scale with a large charge dressing factor. We compare to extremal correlators in superconformal QCD, for which we predict new genus expansions at large charge scaling with $N$.
   
\end{abstract}                                            

\end{center}

\newpage

\pagenumbering{gobble}

\baselineskip .12 in
\setcounter{tocdepth}{2}

\tableofcontents

\setlength{\textheight}{9.0in}

\newpage

\pagenumbering{arabic}
\setcounter{page}{1}

\numberwithin{equation}{section}

\baselineskip=14.8pt
\setcounter{equation}{0}
\setcounter{footnote}{0}

\sec{Introduction}

~~~ What is the large charge limit of semiclassical string theory in AdS? 

In \cite{Paul:2022piq}, we leveraged the S-duality of $SU(N)$ $\cN=4$ super Yang-Mills (SYM) theory and the power of supersymmetric localization to find a recursive expression for a family of integrated four-point functions, indexed by a half-BPS R-charge $p$, that was exact in all parameters: the R-charge $p$, complexified gauge coupling $\t$, and rank $N$. In this paper, we solve this recursion and use it to develop the large charge expansion, $p\gg1$. Expressed in an S-duality-invariant complete eigenbasis, the solution is nothing more than a product of ${}_3F_2$ generalized hypergeometric functions. This allows us to systematically analyze all regimes of large charge $p$, either with fixed $N$ and $\tau$ or in double- or triple-scaling limits. This includes ``gravitational'' double-scaling limits of large $p$, large $N$ with fixed $\a:= p/N^2$ -- a challenging regime for conventional large charge effective field theory techniques -- in which our observable becomes an integrated heavy-heavy-light-light (HHLL) correlator. Holographically, this large charge, large $N$ regime captures semiclassical scattering of type IIB strings in large-momentum deformations of AdS$_5\, \times$ S$^5$: the integrated correlators provide an unusual window into the dynamics of type IIB string theory in this regime, for arbitrary values of the complexified string coupling $\tau = \chi + i e^{-\phi}$.

While the regime of $p \sim N^2$ is perhaps the most novel and physically interesting from the AdS/CFT perspective, the large charge (more generally, large quantum number) expansion of quantum and conformal field theories is a fascinating subject in and of itself; see e.g. \cite{Hellerman:2015nra,Monin:2016jmo,Alvarez-Gaume:2016vff,Hellerman:2017veg,Hellerman:2017sur,Jafferis:2017zna,Hellerman:2018xpi,Hellerman:2020sqj,Hellerman:2021yqz,Hellerman:2021duh,Cuomo:2022kio} for a selection of work and \cite{Gaume:2020bmp} for a recent review. We thoroughly situate our results in this context. Our expansions are developed both with and without double-scaling the Yang-Mills coupling, $g^2$. In the former case, we derive exact results in 't Hooft-type double-scaling limits in which $\l_p := g^2p$ is held fixed at large $p$ and arbitrary finite $N$. These analytic formulas reveal interesting S-duality-invariant functional forms, and non-perturbative corrections interpretable as massive $SL(2,\Z)$ dyons in the Coulomb branch effective field theory. 

Our results may serve as a useful benchmark for further analyses of large charge dynamics in general CFTs, and of strongly coupled string amplitudes in non-vacuum backgrounds.

\vskip .1 in
\centerline{************************}
\vskip .04 in
What quantity are we computing, exactly? Our observable is a family of integrated four-point correlators of composites of ${\bf 20'}$ operators in $SU(N)$ $\mathcal{N}=4$ SYM -- the so-called {\it maximal-trace family} of integrated correlators in \cite{Paul:2022piq}. Let us briefly recall their definition. $\cN=4$ SYM contains half-BPS operators $\cO_p$, Lorentz scalars with conformal dimension $\Delta=p$ that transform in the $[0,p,0]$ representation of the $\mathfrak{su}(4)\simeq \mathfrak{so}(6)$ R-symmetry algebra. The case $p=2$ corresponds to the ${\bf 20'}$, the bottom component of the stress-tensor supermultiplet:
\e{}{\O_2 =\half y_Iy_J\Tr(\Phi^I\Phi^J)\,,}
where $\Phi^I$, with $I = 1,\ldots, 6$, are the six real scalars of the theory and $y_Iy^I=0$. A simple family of higher-charge operators, present for any $N\geq 2$, are the half-BPS multi-trace composites of $\cO_2$:
\begin{align}\label{opdef}
	\O_p := \big[\O_2\big]^{\frac{p}{2}}\,, \quad p\in 2\Z_+\,.
\end{align}
With this identification, the {\it maximal-trace family} of four-point functions was defined in \cite{Paul:2022piq} to be of the type $\langle\O_2\O_2\O_p\O_p\rangle$. Such four-point functions are determined by the constraints of superconformal symmetry up to a single function $\mathcal{H}^{(N)}_p(u,v;\tau)$, where $\t=\theta/2\pi + 4\pi i/g^2$ is the complexified gauge coupling and $(u,v)$ are cross ratios, in which the full $\t$-dependence resides. In \cite{Binder:2019jwn}, the integral
\begin{align}\label{eq:integrated_correlator}
	\mathcal{G}_p^{(N)}(\tau) := -\frac{2}{\pi}\int_0^\infty dr\int_0^\pi d\theta\,\frac{r^3\sin^2\theta}{u^2}\,\mathcal{H}^{(N)}_p(u,r^2;\tau)\,,\quad u = 1+r^2-2r\cos\theta\,,
\end{align}
was found to be given by derivatives of the $\cN=2^*$ partition function with respect to various sources. This powerful connection allows exact computation of these {\it integrated correlators} via the supersymmetric localization of $\cN=2^*$ partition function \cite{Pestun:2007rz, Nekrasov:2002qd} -- which nevertheless depend non-trivially on $\t$. These integrated correlators are usefully viewed as the simplest $\cN=4$ SYM analogue of the extremal two-point functions in $\cN=2$ SCFTs \cite{Baggio:2014ioa,Baggio:2014sna,Baggio:2015vxa,Gerchkovitz:2016gxx,Bourget:2018obm,Beccaria:2018xxl,Grassi:2019txd,Beccaria:2020azj}: both quantities are indexed by R-charge, calculable through localization, and depend non-trivially on $\t$.  

In \cite{Paul:2022piq}, we exploited this connection together with the fact that $\mathcal{G}_p^{(N)}(\t)$ can be very efficiently expressed in the $SL(2,\Z)$ spectral decomposition \cite{Collier:2022emf}. This expands $\mathcal{G}_p^{(N)}(\t)$ in an S-duality-invariant, complete eigenbasis of the scalar Laplacian on $\mathbb{H}^+$:
\begin{align}\label{eq:spectral_decomp}
	\mathcal{G}_p^{(N)}(\tau) = \langle\mathcal{G}_p^{(N)}\rangle +\frac{1}{4\pi i}\int_{\text{Re}\,s=\frac{1}{2}}ds\,\frac{\pi}{\sin(\pi s)}s(1-s)(2s-1)^2g_p^{(N)}(s)\,E^*_s(\tau)\,.
\end{align}
$E^*_s(\tau)$ is the real-analytic (completed) Eisenstein series and $g_p^{(N)}(s)$ are the spectral overlaps, i.e. Petersson inner products, with the Eisenstein series, rescaled by a simple factor.\foot{We recall some properties of the Eisenstein series in Appendix \ref{app:Eisensteins}; see also Section 2 of \cite{Collier:2022emf} for further context.} The first term $\langle\mathcal{G}_p^{(N)}\rangle$ is the average over the $SL(2,\Z)$ fundamental domain -- equivalently, the ensemble average of $\cG_p^{(N)}(\t)$ over the $\cN=4$ conformal manifold -- which admits an explicit expression in terms of harmonic numbers given in \eqref{constantTerm} below. 

The overlaps $g_p^{(N)}(s)$ will be our object of study. Upon trading $\cG_p^{(N)}(\t)$ for $g_p^{(N)}(s)$ by passing to the $SL(2,\Z)$ spectral basis, the utility of this basis is immediately evident: the $g_p^{(N)}(s)$ are simply {\it polynomials} in $s$. These polynomials solve a three-term inhomogeneous recursion in $p$, first found in \cite{Paul:2022piq}, given below in \eqref{eq:recursion_maxtrace}. The mission of this paper is to 
\vskip .05 in

{\bf 1)} Solve the recursion in closed form

{\bf 2)} Expand the solution in all possible regimes of large charge $p \gg 1$

\vskip .05 in
\noindent One of the benefits afforded by our exact solution is the ability to analytically compute large charge limits in which $p$ scales with any power of $N$. Such a regime is not commonly explored in large charge analysis of CFTs, particularly for theories with matrix degrees of freedom.\footnote{Large charge, large $N$ regimes have been previously studied in vector model CFTs \cite{Alvarez-Gaume:2019biu,Giombi:2020enj,Giombi:2022gjj,Dondi:2022zna}.} The double-scaling limit $p\sim N^2$ is where large charge semiclassical string theory enters the picture. We will analytically develop these expansions in all physically distinct regimes of $p, N$ and $\tau$. 

\begin{figure}[t]
\begin{center}
	\hspace{-7.5cm} 
	\begin{minipage}[t]{1.5cm} 
		\centering 
		\includegraphics[scale=0.3]{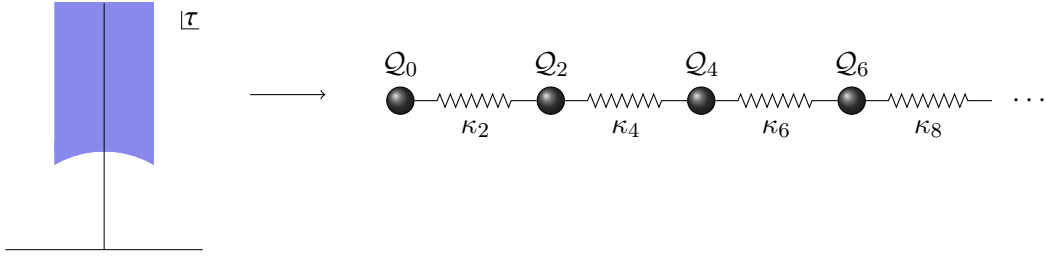}
	\end{minipage} 
	\hspace{1cm} 
		\begin{minipage}[t]{2cm} 
		\centering 
\begin{tikzpicture}[
  mass/.style = {draw,circle,ball color=darkgray},
  spring/.style = {decorate,decoration={zigzag, pre length=.3cm,post length=.3cm,segment length=#1}},
  baseline={([yshift=-12.5ex]current bounding box.center)},vertex/.style={anchor=base,
    circle,fill=black!25,minimum size=18pt,inner sep=2pt}
  ]
  
\draw[black,->] (0,0) -- (1,0);

  \end{tikzpicture}

	\end{minipage} 
	\begin{minipage}[t]{3cm} 
	\centering 

\begin{tikzpicture}[
  mass/.style = {draw,circle,ball color=darkgray},
  spring/.style = {decorate,decoration={zigzag, pre length=.3cm,post length=.3cm,segment length=#1}},
  baseline={([yshift=-12.5ex]current bounding box.center)},vertex/.style={anchor=base,
    circle,fill=black!25,minimum size=18pt,inner sep=2pt}
  ]

  \node[mass,label={above:$\cQ_0$}] (m1) at (0,0) {};
  \node[mass,label={above:$\cQ_2$}] (m2) at (2,0) {};
  \node[mass,label={above:$\cQ_4$}] (m3) at (4,0) {};
  \node[mass,label={above:$\cQ_6$}] (m4) at (6,0) {};
  \node[] (m5) at (8,0) {};
  \node[] (m6) at (10,0) {};

  \draw[spring=4pt] (m1) -- node[below=0.125cm] {$\kappa_2$} (m2);
  \draw[spring=4pt] (m2) -- node[below=0.125cm] {$\kappa_4$} (m3);
  \draw[spring=4pt] (m3) -- node[below=0.125cm] {$\kappa_6$} (m4);
  \draw[spring=4pt] (m4) -- node[below=0.125cm] {$\kappa_8$} (m5);
  \draw (m5) node[right] {$\cdots$} ;
\end{tikzpicture}
\end{minipage} 
	
\vspace{.15in}
\caption{The integrated correlators furnish a semi-infinite lattice of coupled harmonic oscillators. Each oscillator is an integrated correlator $\cQ^{(N)}_p(\t)$, related to $\cG_p^{(N)}(\t)$ by a simple shift defined in \eqref{Qdef}, where $\tau\in\mathcal{F}$, the fundamental domain of $SL(2,\mathbb{Z})$. The oscillators interact with nearest neighbours through a site- and $N$-dependent coupling $\kappa_p$ defined in \eqref{eq:beta}. The set of integrated correlators thus provides a map from the $\cN=4$ conformal manifold to the coupled oscillator system evolving over $\cF$.}
\label{Oscillators}
\end{center}
\end{figure}

In {\bf Section \ref{sec:2}}, we review and solve the recursion relation \eqref{eq:recursion_maxtrace} for the overlaps $g_p^{(N)}(s)$. The solution, given in \eqref{g2Expr}--\eqref{eq:g2_closed}, takes a simple factorised form in terms of generalised hypergeometric functions. This constitutes our first new result. We then provide a physical, and rather striking, picture of the recursion relation: when uplifted to the integrated correlator $\cG_p^{(N)}(\t)$ (and after removing an inhomogeneous piece by a simple shift), it describes a one-dimensional semi-infinite lattice chain of coupled simple harmonic oscillators, evolving on the fundamental domain of $SL(2,\mathbb{Z})$ with nearest-neighbour interactions. This is depicted in Figure \ref{Oscillators}.

In {\bf Section \ref{sec:p>N}}, we consider the large charge $(p\gg1)$ limit with $N$ finite. The spectral overlaps admit an expansion in integer powers of $1/p$. We observe a curious simplification of their analyticity properties for all even $N$, in which $\t$-dependence truncates in the $1/p$ expansion. Within this regime, we may or may not double-scale the Yang-Mills coupling. The limit of fixed $\t$ leads to the expansion \eqref{genericNvsc}, valid for any $N$, with $\tau$-dependence in terms of half-integer Eisenstein series. When specialising to $N=2,3$, we further compute \textit{non-perturbative} corrections to the $1/p$ expansion, which feature the same $SL(2,\mathbb{Z})$-invariant non-holomorphic functions $D_{p/2}(r;\tau)$ recently introduced in \cite{Dorigoni:2022cua}. In our context, these corrections are exponentially small in $\sqrt{p}$, and we identify this precise scale of non-perturbative corrections with the massive propagation of BPS dyonic states on the Coulomb branch, see \eqref{massformula}.

We then analyze the double-scaling limit of fixed $\l_p := g^2 p$. \`A priori, the existence of such a limit is not obvious \cite{Bourget:2018obm,Beccaria:2018xxl,Beccaria:2018owt,Grassi:2019txd,Beccaria:2020azj}, but it follows immediately from the large-$p$ expansion of the spectral overlaps and the properties of the $SL(2,\Z)$ spectral representation. We discuss the leading order `genus-zero' contribution in detail: the exact result, for arbitrary finite $\lambda_p$ and $N$, is given in \eqref{eq:large_p_g0}. (We note that the $N=2$ result in the double-scaling regime may be written in three equivalent forms given in \eqr{eq:large_p_g0_n23}, \eqr{eq:large_p_g0_N=2}, \eqr{eq:strong_coupling_expansion_N2}.)

From there, one can easily extract either the weak or strong coupling expansions in $\lambda_p$. The weak coupling expansion turns out to converge for $|\l_p|< 2\pi^2$. At strong coupling, the leading contribution is \textit{logarithmic} in $\lambda_p$, followed by a (generically) infinite tower of $1/\lambda_p$ corrections. For $N=2,3$, we explicitly compute exponentially small corrections in $\l_p\gg1$. These are controlled by the non-perturbative scale $e^{-\sqrt{2\lambda_p}}$, a result correctly predicted by the {\it weak} coupling radius of convergence as a non-trivial consequence of S-duality \cite{Collier:2022emf}.

In {\bf Section \ref{sec:p~N2}}, we turn to the physically interesting ``gravity regime'' of $p=\a N^2\gg1$ for fixed $\a$. In this limit, the operators $\O_p$ become heavy, dual to a background geometry deformed away from AdS$_5\, \times$ S$^5$ due to the backreaction of a half-BPS gas of $O(N^2)$ supergravitons.\foot{These charged states are heavier than short strings ($\D\sim\l^{1/4}$), semiclassical strings ($\D\sim\sqrt{\l}$), BMN strings ($\D\sim\sqrt{N}$) or D-branes/giant gravitons ($\D\sim N$).}  The integrated correlators are holographically computing integrated string amplitudes in this background in a semiclassical expansion of small $G_N \sim 1/N^2$. Remarkably, we find that the large $N$ expansion \eqref{eq:large_p=N^2} of the spectral overlaps proceeds in \textit{even} powers of $1/N$, as for a bona fide genus expansion. We then analyse the usual 't Hooft limit of fixed $\l:=g^2N$, leading to the new {\it triple-scaled} genus expansion \eqref{eq:triple_scaled_expansion} in which $\a,\l$ are both fixed. We find that the $\l\ll1$ expansion converges for $|\l|<R_\a \pi^2$, where
\begin{align}
	R_\a := 1+2\alpha-2\sqrt{\alpha(\alpha+1)}\,.
\end{align}
We interpret this as a \textit{large charge dressing factor}. This, in turn, has non-trivial implications for the strong coupling expansion at $\l\gg1$: it generates an emergent non-perturbative scale, 
\e{}{e^{-2\sqrt{\lambda/R_\alpha}}\,.}
We suggest a bulk dual interpretation of these exponentially small corrections: they are fundamental string worldsheet instantons, where the on-shell classical string action in the deformed geometry produces the charge-dependent ``screening'' factor $1/\sqrt{R_\a}$ relative to empty AdS$_5\, \times$ S$^5$. Generalizing this result to the limit of fixed $\t$, the non-perturbative scale is likewise interpretable as a rescaled tension of a $(p,q)$-string.

A third and final regime, that of large $p= \a N\gg1$ for fixed $\a$, is discussed in {\bf Section \ref{sec:p~N}}. This ``D-brane regime'' is very similar to the limit of large $N$ with finite $p$, and in fact can be obtained therefrom by subsequently taking $p$ to be of order $N$. As a consequence of large $N$ factorisation, the leading order term is simply the $p=2$ integrated correlator times $\frac{p}{2}$ -- the number of $\O_2$ constituents of $\O_p$. In the 't Hooft limit, this boils down to the equality \eqref{eq:large_p=N_g0}.

In {\bf Section \ref{sec:comparison}}, we provide a comparison of our various large charge results with ``extremal'' chiral primary two-point functions, $ \< O_n(0)\overline{O}_n(\i)\>$, in $\mathcal{N}=2$ superconformal QCD with gauge group $SU(N)$ and $N_f=2N$ flavours. We argue that the natural quantity to compare to our $\cN=4$ SYM integrated correlators is $\mathcal{F}_{2n}^{(N)}(\tau)$, defined in \eqr{fdef} as the logarithm of the extremal two-point function normalised by the corresponding $\mathcal{N}=4$ result. In the large charge regimes where $\mathcal{F}_{2n}^{(N)}(\tau)$ has been studied in the literature, we find many qualitative similarities with our results for $\mathcal{G}_p^{(N)}(\tau)$ upon identifying $p=2n$. Conversely, taking this likeness as a guide, we make precise \textit{predictions} for $\mathcal{F}_{2n}^{(N)}(\tau)$ in regimes \textit{not} studied before, most notably in the gravity regime of $p\sim N^2$. In \eqref{Ftriple}, we claim that $\mathcal{F}_{2n}^{(N)}(\tau)$ admits a genus expansion in the triple-scaled regime of $p=\alpha N^2\gg1$ with $\a$ and $\lambda=g^2 N$ held fixed. This sets a concrete target for future calculations of $\mathcal{N}=2$ SCFT extremal correlators.

In {\bf Section \ref{app:p3}}, we present an analogous exact result for a second infinite class of integrated correlators, namely the \textit{odd-$p$ maximal-trace family}, where instead of $\O_p=\big[\O_2\big]^{p/2}$ we consider $\widetilde\O_p= \O_3\big[\O_2\big]^{(p-3)/2}$ with odd $p\geq3$. This tower of correlators shares many similarities with the even-$p$ case discussed above. In particular, the odd-$p$ maximal-trace overlaps obey a similar recursion relation to \eqref{eq:recursion_maxtrace}, for which we also give a closed-form solution in \eqr{tildegpsol}. We note that in the gravity regime of large $p\sim N^2$, the large charge dressing factor $R_\alpha$ mentioned above is unchanged: it is a universal property of large charge backgrounds created by composites of $\O_2$. We conclude with a conjecture that an analogous coupled harmonic oscillator equation is obeyed by more general families of integrated correlators involving $\O_p^{(m)}=\O_m\big[\O_2\big]^{(p-m)/2}$, for some suitable ``base-operator'' $\O_m$ of finite charge $m$.

Lastly, in the many Appendices \ref{app:Eisensteins}--\ref{besselapp}, we collect further details and derivations.
\vskip .1 in
\noindent {\bf On notation}

\noindent Relative to \cite{Paul:2022piq}, we streamline our notation as follows:
\e{}{\O_p\big|_{\rm here} = \O_p^{(\text{max})}\big|_{\rm there}\,,~~\mathcal{G}_p^{(N)}(\tau)\big|_{\text{here}}=\widehat{\mathcal{G}}_p^{(N|\text{max})}(\tau)\big|_{\text{there}}\,,~~g_p^{(N)}(s)\big|_{\text{here}}=\widehat g_p^{(N|\text{max})}(s)\big|_{\text{there}}\nonumber}

\noindent The hatted notation in \cite{Paul:2022piq} denotes a normalisation by the colour-factor of the two-point function $\langle\O_p^{(\text{max})}\O_p^{(\text{max})}\rangle$, which is given in eq. (A.2) of \cite{Paul:2022piq}. Our conventions otherwise follow \cite{Paul:2022piq}. In these conventions, the relation between $g_2^{(N)}(s)$ here and $f_2^{(N)}(s)$ in \c{Collier:2022emf} is $$ g_2^{(N)}(s)\big|_{\rm here} = \frac{2}{(2s-1)^2(N^2-1)}f_2^{(N)}(s)\big|_{\rm there}~.$$ Finally, we define $\t:=x+iy$, and note that all functions of $\t$ appearing in this work are non-holomorphic.
\vskip .05 in
\noindent {\bf Note added:} Prior to submission, we learned that work on similar topics will appear in \cite{Wen-et-al_first,Wen-et-al_second}.

\sec{Exact solution}\label{sec:2}

We start from the presentation \eqr{eq:spectral_decomp} of the integrated correlator. Our goal is to provide a formula for the spectral overlaps $g_p^{(N)}(s)$, and to understand why the formula is so simple from a physical point of view. 

In \cite{Paul:2022piq} these overlaps were found to obey the following remarkable recursion relation in $p$: 
\begin{align}\label{eq:recursion_maxtrace}
s(1-s)g_{p-2}^{(N)}(s) = -\kappa_p \(g_{p}^{(N)}(s)-g_{p-2}^{(N)}(s)\) + \kappa_{p-2} \(g_{p-2}^{(N)}(s)-g_{p-4}^{(N)}(s)\)+\frac{N^2-1}{2}g_2^{(N)}(s)\,,
\end{align}
with 
\be\label{eq:beta}
\kappa_p := \frac p4\(N^2+p-3\).
\ee
Note the inhomogeneous ``seed'' term proportional to $g_2^{(N)}(s)$, which itself obeys a recursion in $N$\cite{Dorigoni:2021guq,Collier:2022emf}: suppressing the $s$-dependence for clarity,
\e{eq:recursion_g2}{(N+2)N^2(N-1) g_2^{(N+1)}=\left[2(N^2-1)-s(1-s)\right] (N^2-1)g_2^{(N)}-N^2(N+1)(N-2) g_2^{(N-1)}.}
The constant term in \eqr{eq:spectral_decomp}, extracted from the overlap by $\langle\mathcal{G}_p^{(N)}\rangle  = \frac12  g_p^{(N)}(1)$, is
\be\label{constantTerm}
\langle\mathcal{G}_p^{(N)}\rangle = \frac{N(N-1)}{4}\(H_{\frac{1}{2} \left(N^2+p-3\right)} - H_{\frac{1}{2} \left(N^2-3\right)}\),
\ee
where $H_n$ is the $n$'th harmonic number.

These recursion relations can be solved in closed form. As explained in \cite{Paul:2022piq}, the overlaps $g_p^{(N)}(s)$ are proportional to $g_2^{(N)}(s)$,
\begin{align}\label{g2Expr}
	g_p^{(N)}(s) = F_p(N,s)\,g_2^{(N)}(s)\,.
\end{align}
We find the following solutions of \eqref{eq:recursion_maxtrace} and \eqref{eq:recursion_g2}:
\begin{align}\label{eq:Fp_closed}
	\boxed{F_p(N,s) = \frac{N^2-1}{2s(1-s)}\,\[1-{}_3F_2\(-\frac{p}{2},s,1-s;1,\frac{N^2-1}{2};1\)\]}
\end{align}
and\footnote{The result \eqr{eq:g2_closed} is a simpler version of an equivalent expression in eq. (A.29) of \cite{Hatsuda:2022enx}.}
\begin{align}\label{eq:g2_closed}
\boxed{	g_2^{(N)}(s) = \frac{N}{N+1}\,_3F_2\(2-N, s,1-s;3,2;1\)}
\end{align}
Each of these functions is manifestly symmetric under reflection $s\leftrightarrow 1-s$, a necessary property inherited from the reflection symmetry of the completed Eisenstein series. One easily checks that $\langle\mathcal{G}_p^{(N)}\rangle  = \frac12  g_p^{(N)}(1)$ is satisfied \cite{Collier:2022emf} by this solution.

The Eisenstein overlaps \eqr{eq:Fp_closed} and \eqr{eq:g2_closed} furnish the full solution for the integrated correlators for any $p$ and $N$. Their extreme simplicity reinforces the value of the $\sl$ spectral basis.\foot{See \c{Benjamin:2021ygh} for the introduction of this basis in a modern CFT context and \c{Benjamin:2022pnx,Luo:2022tqy,Benjamin:2023nts,Haehl:2023tkr} for other recent works in which this basis has been applied.} For the ``physical'' regime of $N\in \mathbb{Z}_{>1}$ and $p\in 2\mathbb{Z}_+$, these expressions degenerate into even polynomials of $s-{1\o2}$. It is interesting to note that the hypergeometric functions controlling the overlaps are analytic in $s$, and have simple analyticity properties in $p$ and $N$. This gives an analytic continuation of the integrated correlators away from the physical values to continuous $p$ and $N$.\foot{The series representations of the generalised hypergeometric functions appearing in \eqref{eq:Fp_closed} and \eqref{eq:g2_closed} are absolutely convergent for $p+N^2>1$ and $N>-2$, respectively. Within this domain of parameter space, which importantly covers the physically relevant regime $p\geq0$ and $N\geq2$, this defines an \textit{analytic} function in all parameters $p$, $s$ and $N$.}

\ssec{Harmonic oscillators on the $SL(2,\Z)$ fundamental domain}

As explained in \cite{Collier:2022emf,Paul:2022piq}, the polynomiality of $g_p^{(N)}(s)$ for $N\in \mathbb{Z}_{>1}$ and $p\in 2\mathbb{Z}_+$ means that the integrated correlators are actually determined by a finite number of orders in weak coupling perturbation theory. Further evidence for such hidden simplicity comes from the following observation. The recursion formula for the overlaps in \eqref{eq:recursion_maxtrace} is very suggestive of an underlying lattice chain evolving on the fundamental domain of $SL(2,\Z)$. We identify one as follows. Define the shifted quantity
\begin{align}\label{qshift}
q_p^{(N)}(s): =g_p^{(N)}(s)-\frac{N^2-1}{2s(1-s)}\,g_2^{(N)}(s)\,.
\end{align}
This removes the inhomogeneous piece from the recursion \eqref{eq:recursion_maxtrace}, equivalently, the ``1''  of \eqr{eq:Fp_closed}. Lifting the resulting recursion for $g_p^{(N)}(s)$ to a differential equation for the integrated correlator $\cG_p^{(N)}(\t)$ may be done by applying the following map \c{Paul:2022piq}:
\e{stmap}{g_p^{(N)}(s)~~ \mapsto ~~\cG_p^{(N)}(\t) - \< \cG_p^{(N)}\>\,,\qquad s(1-s)~~ \mapsto~~\D_\t\,,}
where $\D_\t$ is the hyperbolic Laplacian. This follows from the spectral representation of $\cG_p^{(N)}(\t)$ and the Laplace equation for the Eisenstein series, $\D_\t E_s^*(\t) = s(1-s)E_s^*(\t)$. Implementing the map \eqr{stmap} with the shift \eqr{qshift}, we find 
\begin{align}\label{eq:lattice_chain}
	\Delta_\tau\,\mathcal{Q}_{p-2}^{(N)}(\tau) &= -\kappa_p\(\mathcal{Q}_{p}^{(N)}(\tau)-\mathcal{Q}_{p-2}^{(N)}(\tau)\)+\kappa_{p-2}\(\mathcal{Q}_{p-2}^{(N)}(\tau)-\mathcal{Q}_{p-4}^{(N)}(\tau)\),
\end{align}
where
\be
\label{Qdef}
\mathcal{Q}_p^{(N)}(\tau) := \mathcal{G}_p^{(N)}(\tau)-\frac{N^2-1}{2}\, \Delta_\t^{-1}\cG_2^{(N)}(\tau)\,.
\ee
Equation \eqref{eq:lattice_chain} describes a one-dimensional semi-infinite lattice of harmonic oscillators $\mathcal{Q}_p^{(N)}(\tau)$ interacting with nearest neighbors, with sites labeled by $p\in 2\Z_{\geq 0}$. The oscillators take values in the $SL(2,\Z)$ fundamental domain. This is depicted in Figure \ref{Oscillators}. The interactions are controlled by a coupling $\kappa_p$, the quantity defined previously in \eqref{eq:beta}. It would be fascinating to understand the fundamental origin of this recasting of the integrated correlators.

\ssec{Toward large charge}

The purpose of the remainder of this paper is to expand our exact, non-perturbative results in various limits of large charge $p$ and large $N$. From the form of \eqr{eq:Fp_closed}, it is manifest that there are three distinct scaling regimes of large $p$: 
\begin{enumerate}[leftmargin=1cm,rightmargin=.7cm,label=\textbf{\arabic*})]
\item {$\bm{p \gg N^2}$}: This regime may be explored with $N$ finite or infinite. (Section \ref{sec:p>N})
\item {$\bm{p \sim N^2}$}: This is the ``gravity regime'' in which the charge scales linearly with the central charge at large $N$. (Section \ref{sec:p~N2})
\item {$\bm{p \ll N^2}$}: This is the regime with fixed $\a := p/N^\g$ with $0 \leq\g<2$.\foot{The limit of large $N$ and {\it fixed} $p$ was studied in Section 6.2 of \cite{Paul:2022piq}.} It includes the physically interesting ``D-brane regime'' of $p \sim N$. (Section \ref{sec:p~N})
\end{enumerate}

The full $N$- and $p$-dependence of the correlator is captured by the overlaps $g_p^{(N)}(s)$, decoupled from the $\t$-dependence by the $SL(2,\Z)$ spectral decomposition. To recover the $\t$-dependence, one simply inserts these overlaps into \eqref{eq:spectral_decomp}. In order to make contact with conventional questions about perturbative expansions and the existence of 't Hooft-type limits, for example, the spectral integral should be manipulated as necessary. In particular, in each of the above regimes, we may wish to further scale $g^2$ with a power of $N$ or $p$. In the ensuing sections we carry out a detailed study of these limits, presenting results both in 't Hooft-type limits of $g^2 \ll 1$ and for fixed $\t$. 

\sec{Large charge, finite $N$}\label{sec:p>N}
We begin by considering the large charge limit with $N$ held fixed:
\e{}{p \gg 1\,,\quad N~\text{fixed}\,.}
This falls into regime {\bf 1)} described above.

Recall that the $p$ dependence of $g_p^{(N)}(s)$ is encoded by $F_p(N,s)$, given in \eqref{eq:Fp_closed}. A simple way to extract the large $p$ behaviour of $F_p(N,s)$ is to use the integral representation of the hypergeometric $_3F_2$, expand the integrand in $p$, and integrate back term-wise. The upshot (see Appendix \ref{app:Fp_expansions} for details) is that $F_p(N,s)$ admits the following large charge expansion:
\begin{align}\label{LargePFiniteNOverlap}
F_p(N, s)=\sum_{g=0}^{\infty} p^{-g}\left[p^{s-1} F^{(g)}(N, s)+p^{-s} F^{(g)}(N, 1-s)\right]-\frac{N^2-1}{2s(s-1)}\,.
\end{align}
The genus-zero overlap is\footnote{We use the term ``genus'' in this context to refer to the order in $1/p$. Even though this is not manifestly a planar expansion in the graphical sense of 't Hooft, we will see in \eqr{LargeChargeGenusExp} that a double-scaling limit of small $g^2$ and large $p$ exists. This 't Hooft-like expansion justifies the use of this terminology, which we then extend in referring to the overlaps themselves.}
\begin{align}\label{LargePFiniteNOverlapg0}
F^{(0)}(N, s) =\frac{2^{s-1} \Gamma\left(\frac{N^2+1}{2}\right) \Gamma\left(s-\frac{1}{2}\right)}{\sqrt{\pi}(s-1) \Gamma(s+1) \Gamma\left(\frac{N^2-3}{2}+s\right)}\,.
\end{align}
Iterating to higher genera yields rational functions of $N$ and $s$ times the genus-zero result:
\begin{align}\label{eq:Fp_higher_g}
\begin{split}
	F^{(1)}(N,s) &= (s-1)\,\frac{N^2-1}{2}\,F^{(0)}(N,s)\,,\\
	F^{(2)}(N,s) &= \frac{(s-2)_2}{12(2 s-3)}\,\Big(3N^4(s-2)-6N^2(s-3)+2 s^2-3 s-12\Big)\,F^{(0)}(N,s)\,,\\
	F^{(3)}(N,s) &= \frac{(s-3)_3}{24 (2 s-3)}\,(N^2-1)\Big(N^4(s-3)-2N^2(s-6)+2 s^2-5 s-9\Big)\,F^{(0)}(N,s)\,.
\end{split}
\end{align}

A few general remarks are in order. Let us first perform a check of the large charge expansion \eqref{LargePFiniteNOverlap}. Evaluating the preceding expressions at $s=1$ and using 
\e{}{\<\cG_p^{(N)}\> = \half F_p(N,1)\,g_2^{(N)}(1)\,,}
yields a match with the large $p$ expansion of the constant term \eqref{constantTerm} appearing in \eqref{eq:spectral_decomp},
\be\label{eq:average_large_p}
\langle\mathcal{G}_p^{(N)}\rangle  = \frac{N(N-1)}{4}\[\log \left(\frac{p}{2}\right)-\psi\left(\frac{N^2-1}{2}\right)-\frac{N^2-2}{p}+\frac{3N^4-12N^2+11}{6p^2}+O(p^{-3})\],
\ee
where $\psi(x)$ is the digamma function. We also elaborate on analytic properties of the large $p$ overlaps $F^{(g)}(N,s)$. From the genus-zero expression \eqref{LargePFiniteNOverlapg0} we note the presence of a pole at $s = 1$  together with a generically infinite tower of poles at 
\e{fppoles}{\qquad\qquad s = -{2k+1\o 2}\,, ~~ k\in \Z_{\geq 0} \qquad \qquad \text{(poles, generic $N$)\,.}}
(The would-be pole at $s=\frac12$ is cancelled by the $(2s-1)^2$ factor in the spectral integral \eqref{eq:spectral_decomp}.) An important observation here is that for even values of $N$, this series of negative half-integers actually truncates, due to the denominator factor $\Gamma\left(\frac{N^2-3}{2}+s\right)$, leaving only the poles at 
\e{eq:truncation_of_poles}{\qquad\qquad s= -\frac12\,,-\frac32\,, \ldots\,, -\frac{N^2-5}{2} \qquad \text{(poles, $N\in 2\Z$)\,.}}
A special case is $N = 2$, for which $F^{(0)}(N,s)$ has {\it no} poles in the region $s<1/2$. Finally, note that $F^{(0)}(N,s)$ has zeroes at negative integers $s=-1,-2,-3,...$ which cancel against the poles coming from $\sin(\pi s)$ in \eqref{eq:spectral_decomp}.\foot{This property is shared by the $p=2$ integrated correlator at genus zero \cite{Collier:2022emf}.} At higher genera, the analytic structure of the overlaps is very similar, being proportional to the genus-zero overlap $F^{(0)}(N,s)$ but lacking the pole at $s=1$.

With the overlaps \eqr{LargePFiniteNOverlapg0} and \eqr{eq:Fp_higher_g} in hand, we now analyze the following two large charge limits at arbitrary finite $N$: 

\begin{enumerate}[leftmargin=1cm,rightmargin=.7cm,label=(\textbf{\roman*})]
\item \textit{Fixed $\t$}: $p\gg1$, $\t~\text{fixed}$
\item \textit{Double-scaled ('t Hooft-like)}: $p\gg 1$, $\l_p:=g^2p$ fixed
\end{enumerate}

The fixed-$\t$ limit exists for any $N$, as the Hilbert space admits representations of unbounded charge. The 't Hooft-like limit has been observed, and derived from a matrix model, in previous studies of extremal two-point functions in $\cN=2$ superconformal QCD (SQCD) \cite{Bourget:2018obm,Beccaria:2018xxl,Beccaria:2018owt,Grassi:2019txd}. For our $\cN=4$ SYM integrated correlators, we prove rather easily that this limit exists. That is, in the double-scaling limit, 
\be\label{LargeChargeGenusExp}
\mathcal{G}_p^{(N)}(\t)= \sum_{\mathfrak{g}=0}^{\infty} p^{-\mathfrak{g}} \,{\mathcal{G}}_{\mathfrak{g}}^{(N )}({\lambda_p}) + O\(e^{-p}\).
\ee 
We have indicated the scale of non-perturbative corrections. This 't Hooft-like structure is manifest in the results \eqref{eq:spectral_decomp} and \eqref{LargePFiniteNOverlap}, as we explain below, whereas the existence of the limit of $\cN=2$ extremal correlators was somewhat hidden \cite{Grassi:2019txd}. We will say more about this comparison to $\cN=2$ SQCD extremal correlators in Section \ref{sec:comparison}. 

\ssec{Expansion at finite coupling and $SL(2,\Z)$ dyons}\label{sec3.1}
We first expand at large $p$ for fixed $\t$. Plugging the large $p$ expansion \eqref{LargePFiniteNOverlap} into the spectral representation \eqref{eq:spectral_decomp} gives 
\begin{empheq}[box=\fbox]{equation}\label{eq:spectral_decomp_largecharge}
\begin{aligned}
\noalign{\vskip4pt}
\mathcal{G}_p^{(N)}(\tau) &= \langle\mathcal{G}_p^{(N)}\rangle + \frac{N^2-1}{2}\, H^{(N)}(\tau)\\
&\quad+\frac{1}{2\pi i}\int_{\text{Re}\,s=\frac{1}{2}}ds\,\frac{\pi}{\sin(\pi s)}s(1-s)(2s-1)^2 \sum_{g=0}^\infty\, p^{s-g-1}\,F^{(g)}(N,s) g_2^{(N)}(s)\,E^*_s(\tau)
\end{aligned}
\end{empheq}
The large $p$ expansion of the average $ \langle\mathcal{G}_p^{(N)}\rangle$ was already noted in \eqref{eq:average_large_p}. The second term in the first line is an interesting $SL(2,\mathbb{Z})$-invariant function which we denote by $H^{(N)}(\tau)$, 
\be\label{HFunction}
H^{(N)}(\tau) := \frac{1}{2\pi i}\int_{\text{Re}\,s=\frac{1}{2}}ds\,\frac{\pi}{\sin(\pi s)}(2s-1)^2g_2^{(N)}(s)\,E^*_s(\tau)\,.
\ee
We say more about this function in Appendix \ref{appendix:H}. As is evident from the above, it is independent of $p$ and therefore contributes only at leading order ($\mathfrak{g}=0$) in the large $p$ expansion. Finally, we turn to the integral in the second line of \eqref{eq:spectral_decomp_largecharge}. To access the perturbative large $p$ expansion, we deform the contour to the left, picking up a pole at $s=0$ from $E^*_s(\t)$ (see \eqr{Eispole}) and the poles \eqr{fppoles}. The pole at $s=0$ contributes
\begin{align}
\begin{split}
	&-\frac{N}{2(N+1)}\sum_{g=0}^\infty p^{-g-1}F^{(g)}(N,0)	\approx -\frac{N(N-1)(N^2-3)}{8}\Bigg(\frac{1}{p}-\frac{N^2-1}{2p^2}+\frac{(N^2-1)(N^2-2)}{3p^3}+\ldots\Bigg),
\end{split}
\end{align}
which we remark is \textit{constant} in $\tau$. The poles \eqr{fppoles} contribute half-integer-index Eisenstein series. Altogether, we find the following expansion to the first few orders in $1/p$, for generic $N$: 
\begin{align}\label{genericNvsc}
\begin{split}
	\mathcal{G}_p^{(N)}(\tau)&=\frac{N(N-1)}{4}\[\log \left(\frac{p}{2}\right)-\psi\left(\frac{N^2-1}{2}\right)\]+\frac{N^2-1}{2}\,H^{(N)}(\tau)+\frac{N(N-1)^2(N+1)}{8}\,{1\o p}\\
	&+\frac{\sqrt{2\pi}\,\Gamma\big(\frac{N^2+1}{2}\big)}{8\,\Gamma\big(\frac{N^2-4}{2}\big)}\,g_2^{(N)}\left(-\tfrac{1}{2}\right)\widetilde{E}_{\frac{3}{2}}(\tau)\,{1\o p^{3\o2}}-\frac{N(N-1)(3N^4-12N^2+13)}{48}\,{1\o p^2}\\
	&-\frac{3\sqrt{2\pi}\,\Gamma\big(\frac{N^2+1}{2}\big)}{32\,\Gamma\big(\frac{N^2-4}{2}\big)}\left[(N^2-1)\,g_2^{(N)}\left(-\tfrac{1}{2}\right)\widetilde{E}_{\frac{3}{2}}(\tau)+\frac{3}{2}(N^2-6)\,g_2^{(N)}\left(-\tfrac{3}{2}\right)\widetilde{E}_{\frac{5}{2}}(\tau)\right]{1\o p^{5\o2}}+O(p^{-3})\,,
\end{split}
\end{align}
with $g_2^{(N)}(s)$ given in \eqref{eq:g2_closed} and we introduced the notation $\widetilde{E}_{s}(\tau):=2\Gamma(s)^{-1} E^*_s(\tau)$. The explicit terms given above reflect the general structure of this expansion: the integer powers in $1/p$ (coming from the ensemble average and the $s=0$ pole) give rise to polynomials in $N$ and no $\tau$-dependence, while half-integer powers in $1/p$ come with more complicated $N$-dependence accompanied by Eisenstein series with half-integer index $s=\frac{3}{2},\frac{5}{2},\ldots,s_*$, where $s_*$ is positively correlated with the power of $1/p$. However, note that for \textit{even} values of $N$ there is an upper bound on $s$, to {\it all} orders in $1/p$, due to the truncation of poles described in equation \eqref{eq:truncation_of_poles}: in that case, the only Eisenstein series appearing have indices $s=\frac{3}{2},\frac{5}{2},\ldots,\frac{N^2-3}{2}$.

Let us give further explicit expressions for the cases $N=2$ and $N=3$, which capture the main features of this expansion and exhibit the curious differences between even and odd $N$. They will also allow us to present some non-perturbative corrections to the $1/p$ expansion.

\sssec*{Example: $\bm{N=2}$}

The genus-zero overlap is
\e{}{F^{(0)}(2,s) = {3\x2^{s-2}\o (s-1)(2s-1)\Gamma(s+1)}\,.}
Higher genera may be read off from \eqr{eq:Fp_higher_g}. These are all regular in the left-half-plane $\Re s < \half$, so only the simple pole at $s=0$ of $E^*_s(\t)$ contributes to the (convergent) perturbative $1/p$ expansion of \eqr{eq:spectral_decomp_largecharge}. This leads to the following large charge expansion:
\begin{align}\label{N2LargeChargeCorr}
\begin{split}
\mathcal{G}_p^{(2)}(\tau) &= {\log p\o2} + \({\log 2+\gamma_E-2\o2 } + \frac{3}{2}H^{(2)}(\t)\)+\frac{3}{4 p}-\frac{13}{24 p^2}+ O(p^{-3}) + \cF^{(2)}_{\text {NP}}(p; \tau)\,,
\end{split}
\end{align}
where $\gamma_E$ is the Euler gamma constant. Note the leading logarithmic behavior. 

We have also indicated the presence of non-perturbative corrections in $p$, denoted by $\cF_{\text {NP}}^{(2)}(p;\t)$. This function must be a modular-invariant function of $\t$. In Appendix \ref{App3} we calculate the leading non-perturbative correction, and find that it takes the following form: 

\be \label{N2LargeChargeCorrNP}
\cF^{(2)}_{\text {NP}}(p; \tau) = \big(2\pi^2 p\big)^{1 / 4} D_{p/ 2}\left(-\frac{1}{4}; \tau\right)+O\(p^{-1 / 4} D_{p/ 2}\({1\o 4};\t\)\),
\ee
where
\begin{align}\label{DFunction}
	D_{p/ 2}(r;\tau):=\sum_{(m,n)\neq(0,0)}\frac{e^{-2\sqrt{2p \,Y_{mn}(\tau)}}}{\big(Y_{mn}(\tau)\big)^r}\,,\qquad Y_{mn}(\tau)=\frac{g^2}{4}\left|m+n\tau\right|^2\,,
\end{align}
is an $SL(2,\mathbb{Z})$-invariant non-holomorphic function introduced in \cite{Dorigoni:2022cua}, whose $SL(2,\Z)$ spectral representation we provide in Appendix \ref{app:D_spectral_rep}. For any fixed $\t$, this function encodes a series of large charge corrections that are exponentially small in $\sqrt{p}$.\footnote{This is in close analogy to the case of the integrated correlator $\cG_2^{(N)}(\t)$ studied in \cite{Dorigoni:2022cua}, where this function appeared in the context of non-perturbative corrections to the large $N$ expansion. A bulk interpretation of exponentially suppressed terms was given in terms of the tension of $(p,q)$ strings \cite{Hatsuda:2022enx, Dorigoni:2022cua}. A generalization of these functions also recently appeared in the study of torodial Casimir energy of the $3d$ critical $O(N)$ vector model at large $N$ \cite{Luo:2022tqy}.} We can give a more physical picture of $D_{p/2}\(-{1\o4};\t\)$ from the Coulomb branch point of view: it is a modular sum over the propagation of $SL(2,\Z)$ dyons, with mass 
\be
\label{massformula}
M = \sqrt{2}\sqrt{p} g |m+n\tau |~.
\ee
We identify the mass scale $M$ with the well-known formula for short massive representations of centrally extended $\cN=4$ supersymmetry algebra \cite{Witten:1978mh, Seiberg:1994rs}: $M=\sqrt{2}|Z|$ where $Z$ is the central charge. Physical states carrying such representations appear in the Coulomb branch vacua where the gauge group is spontaneously broken to $U(1)$ factors. Such dyonic states are characterized by $Z= \sqrt{p} g (m+n\tau )$ where the integers $(m,n)$ are the electric and magnetic charges, respectively, and $\sqrt{p}$ is identified with the Coulomb branch parameter which sets the scale of spontaneous symmetry breaking.\footnote{This $\sqrt{p}$ scaling of the Coulomb branch parameter is in agreement with large charge EFT expectations \cite{Hellerman:2017sur, Grassi:2019txd}.} Subleading terms in \eqr{N2LargeChargeCorrNP} are power-law suppressed relative to the above, but are built from the same class of functions $D_{p/2}(r;\t)$ with greater values of $r$. 

\sssec*{Example: $\bm{N=3}$}

The $N=3$ case is a better representative of the generic case \eqref{genericNvsc} as it receives contributions from both Eisenstein series as well as additional exponentially small terms. The genus-zero overlap is
\begin{align}\label{LargePFiniteNOverlapg0_N=3}
F^{(0)}(3, s) =\frac{3\x 2^{s+2}  \,\Gamma\left(s-\frac{1}{2}\right)}{\sqrt{\pi}(s-1) \Gamma(s+1) \Gamma\left(s+3\right)}\,.
\end{align}
The explicit $1/p$ expansion reads
\begin{align}
\begin{split}
	\mathcal{G}_p^{(3)}(\tau) &= \frac{3}{2} \log p+\left(\frac{6\gamma_{E}-11}{4}-{3\o2}\log 2+4H^{(3)}(\tau)\right)+\frac{6}{p}+\frac{27}{2(2p)^{\frac{3}{2}}}\widetilde{E}_{\frac{3}{2}}(\tau)\\
	&\quad-\frac{37}{2 p^2}-\frac{81}{8(2p)^{\frac{5}{2}}}\left(16\widetilde{E}_{\frac{3}{2}}(\tau)+13\widetilde{E}_{\frac{5}{2}}(\tau)\right)+\frac{84}{p^3}+O(p^{-\frac{7}{2}})\\
	&\quad+\cF^{(3)}_{\text {NP}}(p; \tau)\,.
\end{split}
\end{align}
The $1/p$ expansion in Eisenstein series is asymptotic, as $F^{(0)}(3, s)$ diverges factorially at $s\rar-\i$. The requisite non-perturbative corrections $\cF^{(3)}_{\text {NP}}(p; \tau)$ for fixed $\t$ are again exponentially small in $\sqrt{p}$, involving the same class of functions $D_{p/2}(r;\t)$ found earlier. At leading order, we anticipate\foot{One way to see this is to ``uplift'' the results from the double-scaling limit analyzed in the next section. The idea is to replace the effective 't Hooft coupling $\l_p \mapsto 4\pi p/y$, render the result $SL(2,\Z)$-invariant by performing a Poincar\'e sum over $SL(2,\Z)/\Gamma_\infty$, and use the definition of $D_{p/2}(r;\t)$ as a Poincar\'e sum over the polylogarithm \cite{Dorigoni:2022cua}. Applying this to \eqr{n3resurgence} gives the result quoted below. The claim that this procedure works on the nose for this class observables can be checked by comparing \eqr{N2LargeChargeCorrNP} to \eqr{n2polylog}, which were computed independently, and which match precisely under this map.}
\e{N3LargeChargeCorrNP}{\cF^{(3)}_{\text {NP}}(p; \tau) = c_0 D_{p/2}\left(0; \tau\right)+O\(p^{-1 / 2} D_{p/2}\(\half;\t\)\),}
where the constant $c_0$ may be extracted from \eqr{n3resurgence} together with the median Borel resummation \eqr{median}. This is to be understood in the sense of resummation \cite{Dorigoni:2022cua}. 

\ssec{Double-scaled large charge limit}\label{sec:3.2}

We now study a limit in which a double-scaling 't Hooft-like parameter $\l_p$ is held fixed:
\e{LargeChargetHooft}{ p\gg 1\,,\quad \l_p := g^2p ~ \text{fixed}\,.}
To perform this expansion, we follow the same logic as in \cite{Collier:2022emf,Paul:2022piq}: since all $\t$-dependence sits in the Eisenstein series $E_s^*(\t)$, we simply substitute $y=4\pi/g^{2} = 4\pi p/\lambda_p$ (where $y:= \Im(\t)$). We may then develop the $1/p$ expansion by deforming the spectral contour. In particular, plugging the zero mode $E^*_{s,0}(y) = \L(s) y^s + \L(1-s) y^{1-s}$ and the large $p$ expansion \eqr{LargePFiniteNOverlap} into the spectral decomposition \eqref{eq:spectral_decomp} generates the $1/p$ expansion to all orders:
\begin{align}\label{eq:large_p}
{\mathcal{G}}_p^{(N)}(\tau) &= \langle{\mathcal{G}}_p^{(N)}\rangle + {1\over 2\pi i} \int_{\Re s =\frac{1}{2}} ds \,{\pi\over \sin(\pi s)}(2s-1)^2\Bigg(\frac{N^2-1}{2}g_2^{(N)}(s)\Lambda(s)\,p^{s}\(\frac{\lambda_p}{4\pi}\)^{-s}\\[5pt]
	&\quad+s(1-s)\, g_2^{(N)}(s)\,\sum_{g=0}^{\infty}p^{-g}\[\Lambda(s)\,p^{2s-1}\(\frac{\lambda_p}{4\pi}\)^{-s}+\Lambda(1-s)\(\frac{\lambda_p}{4\pi}\)^{s-1}\]F^{(g)}(N,s)\Bigg)\,,\no
\end{align}
where we recall $\langle{\mathcal{G}}_p^{(N)}\rangle$ in \eqr{constantTerm}, and $\L(s)=\pi^{-s} \Gamma(s)\zeta(2s)$ is the completed Riemann zeta function (see Appendix \ref{app:Eisensteins}). In the second line we used the $s\leftrightarrow 1-s$ symmetry. To develop the large $p$ expansion of the $p^s$ and $p^{2s-1}$ terms we need to close the integration contours to the left. This generates terms with positive powers of $\lambda_p$ (no matter the value of $\lambda_p$). The second term in the second line is exact in $\l_p$, but we may, if desired, close the contour to the left (for $\l_p \gg 1$) or to the right (for $\l_p \ll 1$). 

This establishes the large-charge genus expansion advertised in \eqref{LargeChargeGenusExp}. In the following we work out the exact results at genus zero and genus one.

\subsubsection{The genus-zero correlator}
Computing the $p^0$ contributions from \eqref{eq:large_p} yields
\begin{empheq}[box=\fbox]{equation}\label{eq:large_p_g0}
\begin{aligned}
\noalign{\vskip4pt}
\mathcal{G}_{\mathfrak{g}=0}^{(N)}(\lambda_p) &= \frac{N(N-1)}{4}\[\log \(\frac{\lambda_p}{32 \pi ^2}\)-H_{\frac{N^2-3}{2}}+2\gamma_{E} \]+\frac{N(N+1)}{4}H_N+\frac{N(3N-7)}{8}\\[5pt]
	& + {1\over 2\pi i} \int_{\Re s =\frac{1}{2}} ds \,{\pi\over \sin(\pi s)} s(1-s)(2s-1)^2\,\Lambda(1-s)\(\frac{\lambda_p}{4\pi}\)^{s-1} g_2^{(N)}(s)\, F^{(0)}(N,s)
\end{aligned}
\end{empheq}

\noindent Both lines descends from the respective lines in \eqref{eq:large_p}.\foot{Recalling that the first term in the integrand of \eqref{eq:large_p} has a double pole at $s=0$, the harmonic number $H_N$ comes from the derivative of $g_2^{(N)}(s)$ given in \eqref{g2Expr}, evaluated at $s=0$ -- see \eqr{3F2deriv}.} Note that the appearance of $\log (\lambda_p)$ involves a nice interplay between the $\log (p)$ of \eqref{N2LargeChargeCorr}, on the one hand, and a $\log (g^2)$ term from $H^{(N)}(\t)$, on the other. As derived in \eqr{Hloglp}, these terms add with the same $N$-dependent coefficient, thus giving a $\log (\l_p)$ which is, crucially, consistent with a genus expansion in $1/p$.

This is the full answer for arbitrary finite $\l_p$ and $N$.  We present below the specialized results for $N=2,3$:
\es{eq:large_p_g0_n23}{\mathcal{G}_{\mathfrak{g}=0}^{(2)}(\lambda_p) &= {1\o2}\log\({\l_p\o 8\pi^2}\) + 1+\gamma_E - {1\over 2\pi i} \int_{\Re s =\frac{1}{2}} ds \, (2s-1)\,\Gamma^2(1-s)\,\zeta(2-2s)\(\frac{\lambda_p}{2}\)^{s-1},\\
\mathcal{G}_{\mathfrak{g}=0}^{(3)}(\lambda_p) &= {3\o 2}\log\({\l_p\o 32\pi^2}\) + {7\o2}+ 3\gamma_E  \\~&- {1\over 2\pi i} \int_{\Re s =\frac{1}{2}} ds \, (2s-1)\,\Gamma^2(1-s)\,\zeta(2-2s)\(\frac{\lambda_p}{2}\)^{s-1} \, \(\frac{6(s(s-1)+6)\Gamma \left(s+\frac{1}{2}\right)}{\sqrt{\pi }\Gamma (s+3)}\).}

For completeness, we now study the expansions of \eqr{eq:large_p_g0} at weak and strong coupling.

\subsubsection*{The weak coupling expansion}

Let us start by considering the expansion for $\lambda_p\ll1$, obtained by deforming the integration contour in \eqref{eq:large_p_g0} towards the right. The first pole that one encounters is at $s=1$. This happens to be a double pole, and its contribution cancels against the entire first line of \eqref{eq:large_p_g0}. The remaining poles at integer $s$ are simple poles and we get the following sum over residues:
\begin{align}\label{eq:large_p_g0_weak}
	{\mathcal{G}}_{\mathfrak{g}=0}^{(N)}(\lambda_p) &= -\sum_{s=2}^\infty (-1)^{s} s (s-1)(2s-1)^2 \Lambda\(s-\frac12\) \left(\frac{\lambda _p}{4 \pi }\right)^{s-1} g_2^{(N)}(s) F^{(0)}(N,s)\\
&=	\frac{3 N \zeta (3) \lambda _p}{8 \pi ^2} -\frac{75 N^2 \zeta (5) \lambda _p^2}{128 \pi ^4 \left(N^2+1\right)} + \frac{1225 N^3 \zeta (7) \lambda _p^3}{1024 \pi ^6 (N^2+1)\left(N^2+3\right)} + O(\l_p^4)\,.\no
\end{align}
This is a convergent expansion. The radius of convergence may be straightforwardly derived from the large order asymptotics of the summand, which gives
\be\label{eq:radius}
|\l_p| <2\pi^2\,.
\ee
This is independent of $N$. As explained in \cite{Collier:2022emf}, a consequence of S-duality is that the weak coupling radius of convergence \eqref{eq:radius} directly encodes the scale of exponentially small (non-perturbative) corrections to the \textit{strong} coupling expansion, which we address further below.

Before that, let us derive an alternative integral representation of the genus zero result \eqref{eq:large_p_g0}. Using the zeta function identity
\be\label{eq:zeta_integral_identity}
\zeta(2s-1)=\frac{2^{2s-2}}{\Gamma(2s)}\int_0^{\infty}dw~\frac{w^{2s-1}}{\sinh^2(w)}\,,
\ee
upon commuting the integral and the sum one can actually resum the weak coupling expansion \eqref{eq:large_p_g0_weak} for any given value of $N$. While the generic $N$ expression is not very illuminating, the result for $N=2$ takes the rather simple form\footnote{This expression has been independently obtained by \c{shotaetal} using a complementary method, based on direct analysis in a large charge background of $\cN=4$ SYM. We thank Jo\~ao Caetano, Shota Komatsu and Yifan Wang for discussions.} 

\begin{align}\label{eq:large_p_g0_N=2}
{{\mathcal{G}}_{\mathfrak{g}=0}^{(2)}(\lambda_p)=\int_0^{\infty}dw\,{w\o \sinh^2(w)}\(1-J_0\Big(\tfrac{w\sqrt{2\lambda_p}}{\pi}\Big)\).}
\end{align}

\noindent Note that this representation provides an analytic continuation of the weak coupling expansion, valid beyond the radius of convergence \eqref{eq:radius}, that is fully equivalent to the spectral representation \eqref{eq:large_p_g0_n23} for $N=2$. The same calculation for $N=3$ yields 
\begin{align}\label{eq:large_p_g0_N=3}
\begin{split}
	\mathcal{G}_{\mathfrak{g}=0}^{(3)}(\lambda_p)&=\int_0^{\infty}dw~\frac{3}{w^3\lambda_p^2\sinh^2(w)}\,\bigg(w^4\lambda_p^2+w^2\lambda_p(32\pi^2-w^2\lambda_p)J_0^2\Big(\tfrac{w\sqrt{\lambda_p}}{\sqrt{2}\pi}\Big)\\
	&+8\sqrt{2\lambda_p}w\pi(w^2\lambda_p-16\pi^2)J_0\Big(\tfrac{w\sqrt{\lambda_p}}{\sqrt{2}\pi}\Big)J_1\Big(\tfrac{w\sqrt{\lambda_p}}{\sqrt{2}\pi}\Big)+(w^2\lambda_p-16\pi^2)J_1^2\Big(\tfrac{w\sqrt{\lambda_p}}{\sqrt{2}\pi}\Big)\bigg)\,.
\end{split}
\end{align}

\subsubsection*{The strong coupling expansion and non-perturbative corrections}
To develop the strong coupling expansion, we deform the integration contour of the spectral integral to the left. From the second line of \eqref{eq:large_p_g0} we have 
\begin{align}\label{eq:strong_coupling_expansion}
\begin{split}
\text{Second line of \eqref{eq:large_p_g0}} = \frac{\sqrt{2 \pi } \Gamma \left(\frac{N^2+1}{2}\right)}{\Gamma \left(\frac{N^2-4}{2}\right)}  &\bigg(\frac{2 \zeta (3)}{\lambda_p^{3/2}} g_2^{(N)}\(-\frac12\)-\frac{9 \left(N^2-6\right) \zeta (5)}{\lambda_p^{5/2}} g_2^{(N)}\(-\frac32\)+O(\lambda_p^{-7/2})\bigg).
\end{split}
\end{align}
To all orders this is an expansion in negative half-integer powers of $\lambda_p$ accompanied by single zeta-values times rational functions of $N$. These terms accompany the first line of \eqref{eq:large_p_g0}, which includes the leading $\log(\l_p)$ behavior. As discussed around \eqref{eq:truncation_of_poles}, for even values of $N$ the gamma function prefactor implies that the  (perturbative) strong coupling expansion truncates. An extreme example of this phenomenon is the  $N=2$ case, where the perturbative $1/\lambda_p$ expansion vanishes completely.

Next, let us discuss non-perturbative corrections to the strong coupling expansion \eqref{eq:strong_coupling_expansion}. The presence of these exponentially small corrections, and the precise scale, can be derived from {\it weak} coupling considerations \cite{Collier:2022emf}, in particular from the finite radius of convergence of the weak coupling expansion. This follows from a result proven in Section 7.1 of \cite{Collier:2022emf}.\foot{Let us recall that result. Suppose there exists a double-scaling limit of an $SL(2,\Z)$-invariant gauge theory observable with a coupling (say) $\l$. At any fixed genus, if the $\l\ll1$ expansion is convergent with radius $\l_*$, then subject to a condition on the spectral overlap, the $\l\gg 1$ expansion is non-Borel summable, and requires exponentially small corrections with scale $\L^2(\l) := e^{-2\pi\sqrt{ \l/\l_*}}$. The condition is that the spectral overlap has the same $s\rar\pm \i$ asymptotics up to reflection of $s$. In the context of the $\l_p \gg 1$ expansion, this condition holds for our overlap: the relevant quantity is $g_2^{(N)}(s)\, F^{(0)}(N,s)$, and the result follows because $g_2^{(N)}(s)$ is reflection symmetric, and $F^{(0)}(N,s)$ is a ratio of gamma functions. \label{footref}} Applying that result to the present case, from the weak coupling radius \eqref{eq:radius} we infer the strong coupling non-perturbative scale to be 
\e{npscale}{\L^2(\l_p) := e^{-\sqrt{2\lambda_p}}\,.}
We emphasise that this is independent of $N$. 

It also follows from $SL(2,\Z)$-invariance \cite{Collier:2022emf} that there is (generically) a set of ``S-dual'' corrections controlled by the scale
\e{npscaleS}{\L^2(\l_{p,\mathsf{S}}) := e^{-4\pi p \sqrt{2/\lambda_p}}\,,}
where $\l_{p,\mathsf{S}} := p/g^2 = (4\pi p)^2/\l_p$ is the S-dual coupling to $\l_p$. These come ``for free'' in the S-duality-invariant spectral decomposition, from the reflected branch of the Eisenstein zero mode. Unlike the scale \eqr{npscale}, the scale \eqr{npscaleS} is exponentially small in $p$ at fixed $\l_p$ -- a non-perturbative effect in the charge itself.\foot{In the fixed $\t$ limit studied in the previous section, these terms are part of $D_{p/2}(r;\t)$ \cite{Dorigoni:2022cua}.}

The above results may be confirmed by direct computation for fixed $N$. Due to the different nature of the even versus odd $N$ strong coupling expansions, the computation follows two different approaches which we illustrate by separately considering the cases $N=2$ and $N=3$.

\sssec*{Example: $\bm{N=2}$}

As mentioned before, the perturbative strong coupling expansion \eqref{eq:strong_coupling_expansion} vanishes in this case, and one is left with the constant contributions from the first line of \eqref{eq:large_p_g0_n23}. These will be augmented by exponentially small corrections in $\lambda_p$: such terms can be seen by comparison with the integral representation of ${\mathcal{G}}_{\mathfrak{g}=0}^{(2)}$ derived above by resumming the weak coupling expansion. 

In fact, analysing \eqref{eq:large_p_g0_N=2} numerically, we find that ${\mathcal{G}}_{\mathfrak{g}=0}^{(2)}$ can be written, for {\it finite} $\l_p$, as 
\begin{align}\label{eq:strong_coupling_expansion_N2}
{{\mathcal{G}}_{\mathfrak{g}=0}^{(2)}(\lambda_p) = \frac{1}{2}\log\(\frac{\lambda_p}{8\pi^2}\)+1+\gamma_{E} + \cF^{(2)}_{\text{inst}}(\l_p)}
\end{align}
where $\cF^{(2)}_{\text{inst}}(\l_p)$ is given by the following convergent sum\footnote{This result was inspired by \cite{Grassi:2019txd}, who found a very similar non-perturbative completion of the large charge, double-scaled $\cN=2$ SQCD extremal two-point function. We make further comparisons in Section \ref{sec6.1}.}
\begin{align}\label{n2polylog}
\cF^{(2)}_{\text{inst}}(\l_p) &= \sum_{k=1}^\infty \(k\sqrt{8\lambda_p}  K_1\left(k \sqrt{2\lambda_p}\right)-2 K_0\left(k \sqrt{2\lambda_p}\right)\)
\end{align}
where $K_i(x)$ is the modified Bessel function of the second kind. This may be rewritten in terms of polylogarithms which make the strong coupling expansion manifest:
\es{N2thooftNP}{\cF^{(2)}_{\text{inst}}(\l_p) &= \sqrt{2\pi } {(2\lambda_p)^{1/4} } \text{Li}_{-\frac{1}{2}}\left(e^{-\sqrt{2\lambda_p}}\right)+ O(\lambda_p^{-1/4})}
As a check, this leading term is in precise agreement with the non-perturbative correction obtained from the large charge expansion at finite $\t$ in \eqref{N2LargeChargeCorrNP}. To see this, one compares to the $(m,n)=(\pm 1,0)$ terms in the sum appearing in \eqref{DFunction}, which give the leading contribution in the double scaled limit (see also Appendix \ref{App3}). Note that, using $\Li_{-\half}(x\ll1) \approx x$, the scale of non-perturbative corrections is $e^{-\sqrt{2\lambda_p}}$, consistent with \eqr{npscale}.

\sssec*{Example: $\bm{N=3}$}

For odd values of $N$ the strong coupling expansion does not truncate. For $N=3$,
\begin{align}\label{eq:strong_coupling_expansion_N3}
\begin{split}
{\mathcal{G}}_{\mathfrak{g}=0}^{(3)}(\lambda_p) \approx \frac{3}{2}\log\Big(\frac{\lambda_p}{32\pi^2}\Big)+\frac{7}{2}+3\gamma_{E}+\sum_{n=1}^\infty c_3(n)\,\lambda_p^{-\frac{1}{2}-n}\,,
\end{split}
\end{align}
where
\begin{align}\label{eq:coeffs_c3}
	c_3(n) = 3(4 n^2+23)\,\frac{\Gamma(n+\frac{1}{2})\Gamma(n-\frac{5}{2})\Gamma(2n+1)\zeta(2n+1)}{2^{n-\frac{1}{2}}\pi\Gamma(n)^2}\,.
\end{align}
These coefficients diverge double-factorially and hence the strong coupling expansion requires exponentially small corrections. 
Following \cite{Dorigoni:2021guq}, these corrections can be computed analytically, by applying resurgence techniques to the above asymptotic expansion. The asymptotic expansion is not Borel summable, diverging doubly-factorially without alternating sign, and the necessary non-perturbative corrections are encoded in the discontinuity of the Borel transform (more precisely, the difference between the two lateral resummations above and below the branch cut, which we denote by $\delta\mathcal{G}_{\mathfrak{g}=0}^{(3)}(\lambda_p)$). We defer the details of the computation to Appendix \ref{app:resurgence}, and just quote the result for the discontinuity: 
\begin{align}\label{n3resurgence}
\begin{split}
	\delta\mathcal{G}_{\mathfrak{g}=0}^{(3)}(\lambda_p) = i\,\bigg[&24\,\text{Li}_0\big(e^{-\sqrt{2\lambda_p}}\big)+\frac{102\sqrt{2}}{\lambda_p^{1/2}}\,\text{Li}_1\big(e^{-\sqrt{2\lambda_p}}\big)+\frac{1839}{2\lambda_p}\,\text{Li}_2\big(e^{-\sqrt{2\lambda_p}}\big)\\
	&+\frac{20211}{4\sqrt{2}\lambda_p^{3/2}}\,\text{Li}_3\big(e^{-\sqrt{2\lambda_p}}\big)+\frac{578817}{64\lambda_p^2}\,\text{Li}_4\big(e^{-\sqrt{2\lambda_p}}\big)+\ldots\bigg].
\end{split}
\end{align}
This is an expansion in the correct non-perturbative scale \eqr{npscale}, obtained by a completely different method as the $N=2$ case. A similar calculation may be done to find the exponentially suppressed terms \eqr{npscaleS}; see \cite{Collier:2022emf} for an analogous calculation in the large $N$ 't Hooft limit of $\cG_2^{(N)}(\t)$.

\subsubsection{Higher-genus correlators}

An analogous analysis as performed for the genus-zero correlator above can be straightforwardly carried out for higher genera. The main difference is that at strong coupling $\l_p\gg1$, there are no $\log(\lambda_p)$ contributions; instead, a finite number of terms with positive powers of $\lambda_p$ will appear. Note that the weak coupling radius of convergence will always be given by $2\pi^2$,\footnote{This follows from the form of the higher-genus overlaps $F^{(g)}(N,s)$: as seen in \eqref{eq:Fp_higher_g}, the $F^{(g)}(N,s)$ are related to the genus-zero result $F^{(0)}(N,s)$ by a rational function in $s$ which does not change their large $s$ asymptotics.} and therefore the scale of non-perturbative corrections, $e^{-\sqrt{2\lambda_p}}$, is the same for all $\mathfrak{g}$.

For instance, the genus-one correlator reads
\begin{empheq}[box=\fbox]{equation}\label{eq:large_p_g1}
\begin{aligned}
\noalign{\vskip4pt}
	\mathcal{G}_{\mathfrak{g}=1}^{(N)}(\lambda_p) &=-\frac{3N(N^2-1)\zeta(3)}{16\pi^2}\,\lambda_p +\frac{N(N^2-1)(N-1)}{8}\\
	& + {1\over 2\pi i} \int_{\Re s =\frac{1}{2}} ds \,{\pi\over \sin(\pi s)} s(1-s)(2s-1)^2\,\Lambda(1-s)\(\frac{\lambda_p}{4\pi}\)^{s-1}g_2^{(N)}(s)\,F^{(1)}(N,s)
\end{aligned}
\end{empheq}
Following the same steps described above, one may develop the weak and strong coupling expansions, compute the exponentially small corrections at large $\lambda_p$, and derive Bessel representations. Some explicit results for the $\mathfrak{g}=1$ term are collected in Appendix \ref{app:large_p_g1}. Here we simply give the Bessel representation for the $N=2$ case:
\begin{align}\label{eq:large_p_g1_N=2}
	\mathcal{G}_{\mathfrak{g}=1}^{(2)}(\lambda_p)=-\int_0^{\infty}dw~\frac{3w^2}{4\pi^2\sinh^2(w)}~\bigg(w\lambda_p-\sqrt{2\lambda_p}\pi J_1\Big(\tfrac{w\sqrt{2\lambda_p}}{\pi}\Big)\bigg).
\end{align}

\sec{Large charge, $p=\a N^2$}\label{sec:p~N2}
In this section, we discuss the physically interesting `gravity regime', in which the charge $p$ scales linearly with the central charge, 
\e{}{ p\gg 1\,,\quad N \gg1\,,\quad \alpha={p\o N^2}~ \text{fixed}\,,\quad \a\in \R_{+}\,.}
We recall that the central charge $c \sim N^2/4$ to leading order in large $N$. 

In this regime, $\cO_p$ is ``heavy,'' with $\D = p \sim c$. Consequently, we may think of the  four-point function in vacuum as a two-point function $\<p|\cO_2\cO_2|p\>$ in a heavy state $|p\>:= \cO_p(0)|0\>$ defined in radial quantization:
\e{}{\langle\O_p(0)\O_2(z,\zb)\O_2(1)\O_p(\i)\rangle = \<p|\cO_2(z,\zb)\cO_2(1)|p\>\,.}
This is a heavy-heavy-light-light (HHLL) correlator, more general versions of which have been studied in various approximate regimes in $d\geq2$ holographic CFTs  \c{Fitzpatrick:2015zha,Giusto:2018ovt,Karlsson:2019dbd,Kulaxizi:2019tkd,Fitzpatrick:2020yjb,Li:2020dqm,Parnachev:2020fna,Giusto:2020mup, Karlsson:2021duj,Dodelson:2022eiz}.\foot{In explicit computations of HHLL correlators in $d>2$ CFTs with Einstein gravity duals, the limit $\a \ll 1$ is taken, for technical reasons. In contrast, in our supersymmetric and integrated setup, we can compute for finite $\a$. It is of clear interest whether the results of this paper can inform finite-$\alpha$ expressions for HHLL correlators in holographic CFTs.}  

In the gravity dual, this is a large charge regime of semiclassical string theory. The geometry dual to $|p\>$ is deformed away from AdS$_5\, \times$ S$^5$ due to large backreaction, with asymptotic charges matching those of the half-BPS state $|p\>$. Since $\O_p$ is a composite of $\O_2$'s, the dual geometry may be thought of as generated by the classical gravitational backreaction of a gas of $p/2 \sim O(1/G_N)$ supergravitons. While not a black hole (which do not exist for half-BPS states \cite{Gutowski:2004ez,Gutowski:2004yv}), and not quite an LLM geometry, the half-BPS condition suggests that it should be possible to construct the bulk dual to this state.\footnote{In \cite{Skenderis:2007yb}, a precise coherent state which \textit{is} dual to a LLM geometry has been constructed at large $N$, and consists of an \textit{infinite sum} of states $|p\>$ (not all of which are heavy). We thank Iosif Bena, Shota Komatsu, Kostas Skenderis and David Turton for discussions on this point.} Thanks to the exact solution of the spectral overlaps in \eqref{eq:Fp_closed} and \eqref{eq:g2_closed}, we can access this non-trivial regime for arbitrary $\t$ from the $\cN=4$ SYM side. This is a prediction for a bulk string theory computation at finite axio-dilaton $\t = \chi + ie^{-\phi}$. 

We now present this solution. As explained in Appendix \ref{app:Fp_expansions}, when $p=\alpha N^2$ the function $F_p(N,s)$ organises into an expansion in \textit{even} powers of $1/N^2$. The same holds for the $p=2$ overlaps $g_2^{(N)}(s)$ in the large $N$ limit \cite{Collier:2022emf}, such that the maximal-trace overlaps $g_p^{(N)}(s)$ admit an expansion of the form 
\begin{align}\label{eq:large_p=N^2}
	g_{\alpha N^2}^{(N)}(s)=\sum_{g=0}^\infty N^{2-2g}\big[N^{s-1}h_\alpha^{(g)}(s)+N^{-s}h_\alpha^{(g)}(1-s)\big]\,.
\end{align}
As suggested by the summation index $g$, we indeed interpret this as a genuine {\it genus expansion}, which becomes manifest in \eqr{eq:triple_scaled_expansion}. Though it is suggested by thinking about the string theory side, the existence of such an expansion is non-trivial from the field theory point of view: $\O_p$ is a heavy operator, so planar combinatorics do not {\it a priori} apply to $\<p|\cO_2\cO_2|p\>$.

The explicit form of the genus-zero overlap $h_\alpha^{(0)}(s)$ is obtained from \eqref{eq:c_alpha} together with \eqref{higherggrav} and yields
\begin{align}\label{eq:h0}
	\boxed{h_{\alpha}^{(0)}(s) = \frac{1-\,_2F_1(s,1-s;1;-\alpha)}{2s(1-s)}\,g_2^{(0)}(s)}
\end{align}
where $g_2^{(0)}(s)$ denotes the genus-zero $p=2$ overlap which for convenience we recall here:
\begin{align}\label{eq:g2_genus0}
	g_2^{(0)}(s)=\frac{2^{2 s} \Gamma \left(s+\frac{1}{2}\right)}{\sqrt{\pi } (2 s-1) \Gamma (s+1) \Gamma (s+2)}\,.
\end{align}
This is the leading-order result in the gravity regime.\foot{At risk of repetition, we remind the reader that one may restore the $\t$-dependence by plugging \eqr{eq:large_p=N^2}--\eqr{eq:g2_genus0} into the integral \eqr{eq:spectral_decomp}.} Given the complexity of this regime -- large $N$ and large charge $p \sim 1/G_N$, for arbitrary coupling $\t$ -- this result involves remarkably simple functions. This simplicity persists to higher genus, where such terms, given in Appendix \ref{app:assembly}, probe loop-level effects in semiclassical string theory.

Some comments are in order. First, note that the apparent poles at $s=0,1$ are cancelled by zeroes in the numerator. As such, $h_\alpha^{(0)}(s)$ inherits the familiar pole structure from $g_2^{(0)}(s)$, i.e. it has an infinite tower of simple poles at negative half-integers. On the other hand, the presence of the hypergeometric function in the numerator of \eqref{eq:h0} changes the asymptotic behaviour at large $s$. As we will discuss shortly, one consequence of this is the appearance of a new scale of non-perturbative corrections to the strong coupling expansion. One notes further that evaluating the hypergeometric function at $\alpha=0$ yields the `1' in the numerator of \eqref{eq:h0}, and hence the genus-zero overlap $h_\alpha^{(0)}(s)$ naturally splits into two parts: 
\begin{align}\label{eq:h0_split}
	h_\alpha^{(0)}(s)=\widetilde h_0(s)-\widetilde h_\alpha(s)\,, \qquad \text{with }~\widetilde h_\alpha(s) = \frac{_2F_1(1-s,s;1;-\alpha)}{2s(1-s)}\,g_2^{(0)}(s)\,.
\end{align}
We note in passing that the first term $\widetilde h_0(s)$ is exactly one half times the spectral overlap of the function $H^{(N)}(\tau)$ at leading order at large $N$, c.f. the definition \eqref{HFunction}.

To perform a consistency check of the large $N$ expansion \eqref{eq:large_p=N^2}, we can consider the corresponding expansion of the ensemble average \eqref{constantTerm} for which one has
\begin{align}\label{eq:average_p=N^2}
	\langle\mathcal{G}_{\alpha N^2}^{(N)}\rangle=\frac{N(N-1)}{4}\bigg[\log(\alpha+1)+\frac{2 \alpha}{(\alpha+1)N^2}+\frac{11\alpha(\alpha+2)}{6(\alpha+1)^2 N^4}+O(N^{-6})\bigg].
\end{align}
Evaluating \eqref{eq:large_p=N^2} at $s=1$ and comparing term by term to the above expansion, we indeed find the correct relation $\langle\mathcal{G}_{\alpha N^2}^{(N)}\rangle=\frac{1}{2}g_{\alpha N^2}^{(N)}(1)$.

There are various large $N$ limits to analyse in this regime of $p=\alpha N^2 \gg 1$:

\begin{enumerate}[leftmargin=1cm,rightmargin=.7cm,label=(\textbf{\roman*})]
\item \textit{Fixed $\tau$}
\item \textit{'t Hooft}: $\lambda=g^2N$ fixed 
\item \textit{'t Hooft-like}: $\lambda_p=g^2p$ fixed
\end{enumerate}
We will mainly discuss the 't Hooft limit, a physically interesting case which furthermore connects directly onto supergravity. We make a short comment on the fixed-$\t$ limit at the end of the section. The remaining limits may be developed in direct parallel with previous sections of this paper.

\subsection{'t Hooft limit}\label{sec:tHooft_N}

The 't Hooft limit is obtained by plugging the double-scaled large $N$ expansion of the spectral overlaps \eqref{eq:large_p=N^2} together with the zero-mode of the Eisenstein series into the spectral integral \eqref{eq:spectral_decomp}. This leads to a genus expansion,
\begin{align}\label{eq:triple_scaled_expansion}
{	\mathcal{G}_{\alpha N^2}^{(N)}(\lambda) = \sum_{\mathfrak{g}=0}^{\infty} N^{2-2\mathfrak{g}}\,\mathcal{G}_\alpha^{(\mathfrak{g})}(\lambda) + (\text{non-perturbative})}\,.
\end{align}
The leading order result is
\begin{align}\label{eq:genus0_p=N^2}
\boxed{\mathcal{G}_{\alpha}^{(0)}(\lambda) = \frac{\log(\alpha+1)}{4} + {1\over 2\pi i} \int_{\Re s =\frac{1}{2}} ds \,{\pi\over \sin(\pi s)}\,s(1-s)(2s-1)^2\Lambda(1-s)\left(\frac{\lambda}{4\pi}\right)^{s-1}h_\alpha^{(0)}(s)}
\end{align}
with $h_\alpha^{(0)}(s)$ given in \eqr{eq:h0}. 
On the bulk side, this is the classical string theory result for finite $\a'$. Higher-genus terms, dual to integrated string loop amplitudes, may be likewise assembled using the overlaps $h^{(g)}_\a(s)$ in Appendix \ref{app:assembly}. 

A very interesting field theory challenge for the future is to derive \eqr{eq:genus0_p=N^2} from a large charge effective field theory perspective, in which $N, p$ and $g^2$ are scaled simultaneously with fixed parameters $\l$ and $\a$. What is the corresponding ``triple-scaled matrix model''?

As usual, from the spectral representation we can develop the (perturbative) weak and strong coupling expansions in $\lambda$ by deforming the contour to the right or left, respectively.

\subsubsection{Weak coupling}
Deforming the contour in \eqref{eq:genus0_p=N^2} to the right and summing over residues yields
\begin{align}\label{eq:large_p=N^2_weak}
	\mathcal{G}_\alpha^{(0)}(\lambda) = -\sum_{s=2}^\infty (-1)^{s} s (1-s)(2s-1)^2 \Lambda\(s-\frac12\) \left(\frac{\lambda}{4 \pi }\right)^{s-1} h_\alpha^{(0)}(s)\,,
\end{align}
which for the first few orders evaluates to
\begin{align}
	\mathcal{G}_\alpha^{(0)}(\lambda) \approx \frac{3 \alpha \zeta (3)}{8 \pi ^2}\,\lambda -\frac{75 \alpha  (\alpha +1) \zeta (5)}{128 \pi ^4}\,\lambda^2+\frac{245 \alpha (10 \alpha ^2+15 \alpha +6) \zeta (7)}{2048 \pi ^6}\,\lambda ^3+O(\lambda^4)\,.
\end{align}

More interestingly, we can compute the radius of convergence of the above weak coupling expansion. To this end it is instructive to split the infinite sum \eqref{eq:large_p=N^2_weak} according to the natural splitting of $h_\alpha^{(0)}(s)$ observed in \eqref{eq:h0_split}. As we will see momentarily, this is justified since both sums have a finite radius of convergence. 

For the first sum over $\widetilde h_0(s)$, we find the canonical radius of convergence,
\begin{align}\label{eq:R1}
	R_1 = \pi^2\,.
\end{align}

For the second sum over $\widetilde h_\alpha(s)$ with non-zero $\alpha$, we need to understand the large $s$ asymptotic behaviour of $\widetilde h_\alpha(s)$. It is useful to note that the hypergeometric function can be expressed as a Legendre polynomial: $\,_2F_1(1-s,s;1;-\alpha)=P_{s-1}(2\alpha+1)$. Then, using the expansion of the Legendre polynomial for large index, we find to leading order at large $s$
\begin{align}
	_2F_1(1-s,s;1;-\alpha) \approx \frac{1}{\big(\alpha(\alpha+1)\big)^{\frac{1}{4}}}~\frac{e^{(s-\frac{1}{2})\cosh^{-1}(2\alpha+1)}}{\sqrt{2\pi(2s-1)}}+\ldots.
\end{align}
With this in hand, the radius of convergence of the second sum is given by 
\e{eq:R2}{R_2 = R_\a \pi^2\,,}
where $R_\a$ is a {\it large charge dressing factor}:
\begin{empheq}[box=\fbox]{equation}\label{eq:Ralpha}
\begin{aligned}
\noalign{\vskip4pt}
	R_\a := 1+2\alpha-2\sqrt{\alpha(\alpha +1)} =  e^{-\cosh^{-1}(2\a+1)}
\end{aligned}
\end{empheq}
The larger the charge $\a = p/N^2$, the smaller the radius $R_\a$. For $\a\geq 0$, we find $R_\a \in (0,1]$, saturating the upper bound at $\a=0$ and decreasing monotonically with $\a$. It follows that the total radius of convergence of the combined sum \eqref{eq:large_p=N^2_weak} is given by the smaller of the two, namely, $R_2$.\foot{One can check that the higher-genus weak coupling expansions have the same radius of convergence $R_2$.} 

The dressing factor $R_\a$ may presumably be interpreted as an emergent scale in the large charge EFT, for instance, as an EFT saddle point in the semiclassical limit $p = \a N^2$ being considered here. We will address its meaning in holography shortly.

Let us also quote the Bessel integral representation for $\mathcal{G}_\alpha^{(0)}(\lambda)$, derived in Appendix \ref{besselapp}:
\begin{align}\label{besselrepfinal}
	\mathcal{G}_\alpha^{(0)}(\lambda) = {2\pi^2\o\lambda}\int_0^\infty dw~{K_{11}\({w \sqrt{\lambda}\o\pi};0\) - K_{11}\({w \sqrt{\lambda}\o\pi};\a\)\o w \sinh^2(w)}\,,
\end{align}
where we have defined a localized integral Bessel kernel,
\e{Kdef}{K_{11}(x;\a) := \frac{1}{\pi}\int_0^\pi dt~\frac{J_1(x \sqrt{u_\a})^2}{u_\a}\,,}
where $u_\a:=1+2\a-2\sqrt{\alpha(\alpha+1)}\cos(t)$. Note that $K_{11}\({w \sqrt{\lambda}\o\pi};0\) = J_1\({w \sqrt{\lambda}\o\pi}\)^2$.

\subsubsection{Strong coupling and a new non-perturbative scale}

The strong coupling expansion of $\mathcal{G}_\alpha^{(0)}(\lambda)$ is obtained by deforming the contour in \eqref{eq:genus0_p=N^2} to the left. Picking the poles from $g_2^{(0)}(s)$ at $s=-\frac{1}{2},-\frac{3}{2},\ldots$, for $\lambda\gg1$ one finds
\es{stronggrav}{	\mathcal{G}_\alpha^{(0)}(\lambda)&\approx \frac{\log(\alpha+1)}{4} + 4\Big(1-\,_2F_1\big(-\tfrac{1}{2},\tfrac{3}{2};1;-\alpha\big)\Big)\,\frac{\zeta(3)}{\lambda^{3/2}} - 3\Big(1-\,_2F_1\big(-\tfrac{3}{2},\tfrac{5}{2};1;-\alpha \big)\Big)\,\frac{\zeta(5) }{\lambda^{5/2}}\\
	&\quad+O(\lambda^{-7/2})\,.}
The leading term makes a prediction for the supergravity result:
\e{}{\mathcal{G}_\alpha^{(\rm sugra)} = \frac{\log(\alpha+1)}{4}\,.}
This should be computable from the supergraviton two-point function in the bulk geometry dual to $|p\>$, integrated over boundary positions with the correct measure.

The expansion \eqr{stronggrav} is asymptotic and requires exponentially small corrections. As explained in footnote \ref{footref}, the scale of such corrections is determined by the weak coupling radius of convergence. In the present case, this leads to \textit{two different} sets of non-perturbative corrections: due to the splitting of the genus-zero overlap into $h_\alpha^{(0)}(s)=\widetilde{h}_0(s)-\widetilde{h}_\alpha(s)$, there will be corrections with the usual scale obtained from \eqref{eq:R1},
\begin{align}
	\widetilde{h}_0(s): \qquad \L^2(\l) := e^{-2\sqrt{\lambda}}\,,
\end{align}
together with another set of corrections controlled by a new, $\alpha$-dependent scale obtained from \eqref{eq:R2},
\begin{align}\label{eq:scale_new}
	\widetilde{h}_\alpha(s): \qquad \L^2_\a(\l) := e^{-2\sqrt{\lambda/R_\alpha}}\,,
\end{align}
where we recall the definition $R_\a=1+2\alpha-2\sqrt{\alpha(\alpha +1)}$ from \eqref{eq:Ralpha}.

The above result makes a prediction for a new class of non-perturbative corrections in the gravity regime of large charge. More specifically, these are corrections in the $\l\gg1$ regime, so this is a prediction for a non-perturbative effect in AdS$_5\, \times$ S$^5$ {\it supergravity}. In many instances of AdS$_5\, \times$ S$^5$ holography, terms controlled by $e^{-2\sqrt{\l}}$ are dual to fundamental string worldsheet instantons, where
\e{}{2\pi T_{\rm F1} = \sqrt{\l}\,,}
is the tension of a fundamental string in AdS units. For integrated correlators, such corrections would presumably be captured by a fundamental string worldsheet action of an appropriate bulk configuration (though no such bulk calculation has been done, either for $\cG_2^{(N)}(\t)$ or a higher-charge correlator). This raises the obvious question: what is the picture for the new scale $e^{-2\sqrt{\lambda/R_\alpha}}$ derived above? 

Recall that we are computing integrated two-point functions in large charge backgrounds, schematically $\int \<p|22|p\>$. Therefore, one expects that the above exponential scales may be derived from a fundamental string worldsheet action in the appropriate supergravity background dual to the state $|p\>$. The large charge dressing factor appears from the worldsheet action evaluated on an appropriate cycle in this background, generating a $1/\sqrt{R_\a}$. Since $R_\a\leq 1$, the background charge increases the area relative to empty AdS$_5\, \times$ S$^5$, which has $\a=0$ and $R_0=1$. 

These non-perturbative terms generalize directly to the fixed-$\tau$ limit. Rendering the above argument $SL(2,\Z)$-invariant, non-perturbative corrections should take the form of a modular sum over charge-dependent dyons labeled by $(m,n)\neq (0,0)$, with a rescaling by the dressing factor. The dyonic scale is
\be
\label{massformula2}
\L^2_{m,n; \a} := e^{-{M/ \sqrt{R_\a}}}\,,\quad \text{where}~~M = 2\sqrt{N} g |m+n\tau |~.
\ee
This is suggested by the bulk interpretation: viewing each term in the sum over $(m,n)$ as an $(m,n)$-string probing the deformed geometry, the worldsheet area is independent of the string charges. 

Finding these exponential corrections with the simple factor $1/\sqrt{R_\a}$ explicitly on the string theory side would be a worthwhile computation. Likewise, one should seek these terms as exponentially small corrections to the large charge EFT from massive particle propagation in this novel regime $p =\a N^2 \gg1$, generalizing the analyses of \c{Hellerman:2021yqz,Hellerman:2021duh}.

\sec{Large charge, $p=\alpha N$}\label{sec:p~N}

In this section, we consider the double-scaling limit of 
\e{}{ p\gg 1\,,\quad N \gg1\,,\quad \alpha={p\o N}~ \text{fixed}\,,\quad \a\in \R_{+}\,.}
As $p$ scales linearly with $N$, we will use the terms ``large charge'' and ``large $N$'' interchangeably. 

Due to the structure of the function $F_p(N,s)$ in \eqr{eq:Fp_closed}, the results in this section generalize directly to any double-scaling regime $p=\alpha N^\gamma$ with $0<\gamma<2$. We have chosen to display $\g=1$ because the regime of operator dimensions $\D \sim N$ is physically interesting from the holographic point of view: in particular, $\D \sim N$ states are believed to be dual to giant graviton/D-brane states in AdS$_5\, \times$ S$^5$. Moreover, when $p \sim N$, the 't Hooft limit of fixed $\l = g^2 N$ is equivalent to fixed $\l_p = g^2 p$ up to an order one constant:
\e{}{\l_p = \a \l\,.}

This regime is deceptively simple: from the exact solution \eqr{eq:Fp_closed}, we know that for $p \ll N^2$, the behavior is qualitatively the same as the $1 \ll p \ll N^2$ expansion.  Indeed, as shown in more detail in Appendix \ref{app:Fp_expansions} (see the discussion around equation \eqref{eq:F_large_p=N}), starting from the $1/N$ expansion of $F_p(N,s)$ for finite $p$, and subsequently taking $p$ to be of order $N$, leads to a reshuffling of only a \textit{finite} number of terms in the large $N$ expansion of $F_p(N,s)$. Furthermore, the {\it leading order} term is just the $p=2$ result rescaled by a constant, because of large $N$ factorisation. 

More explicitly, one finds that in the $p=\alpha N$ double-scaling limit the spectral overlaps $g_p^{(N)}(s)$ have a large $N$ expansion of the form 
\begin{align}\label{eq:large_p=N_overlaps}
	g_{\alpha N}^{(N)}(s)=\sum_{g=0}^\infty N^{-g}\big[N^{s}k_\alpha^{(g)}(s)+N^{1-s}k_\alpha^{(g)}(1-s)\big]\,.
\end{align}
The leading order term is simply given by
\begin{align}\label{eq:k0}
	k^{(0)}_\alpha(s)=\frac{\alpha}{2}\,g_2^{(0)}(s)\,,
\end{align}
where the familiar genus-zero $p=2$ overlap $g_2^{(0)}(s)$ has been quoted previously in \eqref{eq:g2_genus0}. Compared to the large $N$ limit with $p$ finite, the above expansion proceeds in powers of $1/N$ rather than $1/N^2$. Further subleading terms can be obtained from the expansion described in Appendix \ref{app:Fp_expansions}, see e.g. equation \eqref{eq:F_large_p=N}. For instance, the first subleading term reads
\begin{align}\label{eq:k1}
\begin{split}
	k^{(1)}_\alpha(s)&=\frac{\alpha^2 (s-2) (s+1)}{8}\,g_2^{(0)}(s)\,.
\end{split}
\end{align}
The other relevant quantity is the ensemble average $\langle\mathcal{G}_p^{(N)}\rangle$ given in \eqref{constantTerm}. In the regime $p=\alpha N$ it has the expansion 
\begin{align}\label{eq:average_p=N}
\begin{split}
	\langle\mathcal{G}_{\alpha N}^{(N)}\rangle &=\alpha\,\bigg[\frac{N}{4}-\frac{\alpha +2}{8}+\frac{2 \alpha ^2+3 \alpha +12}{24N}-\frac{3 \alpha ^3+4 \alpha ^2+24 \alpha +24}{48 N^2}\\
	&\qquad~+\frac{12 \alpha ^4+15 \alpha ^3+120 \alpha ^2+120 \alpha +220}{240N^3}+O(N^{-4})\bigg]\,.
\end{split}
\end{align}

\ssec{'t Hooft limit}
To illustrate some analogies with the large $N$ expansion of the $p=2$ integrated correlator $\mathcal{G}_2^{(N)}(\tau)$, let us consider the 't Hooft expansion of fixed $\l = \l_p/\a$. 

Plugging in the large $N$ expansion \eqref{eq:large_p=N_overlaps} into the spectral decomposition \eqref{eq:spectral_decomp}, one finds that the correlator organises into an expansion of the form %
\begin{align}\label{eq:large_p=N}
{	\mathcal{G}_{\alpha N}^{(N)}(\lambda) = \sum_{\mathfrak{g}=0}^\i N^{1-\mathfrak{g}}\, \mathcal{K}_{\alpha}^{(\mathfrak{g})}(\lambda) + (\text{non-perturbative in $N$})}
\end{align}

Focussing first on the leading order $\mathfrak{g}=0$ contribution, we find as a consequence of equation \eqref{eq:k0} and after replacing $\alpha=p/N$ that
\begin{align}\label{eq:large_p=N_g0}
	\mathcal{K}_{\alpha}^{(0)}(\lambda) = \frac{p}{2}\,\mathcal{G}_{2}^{(\mathfrak{g}=0)}(\lambda)\,,
\end{align}
where the $p=2$ genus-zero term $\mathcal{G}_{2}^{(\mathfrak{g}=0)}(\lambda)$ has been studied previously in \cite{Binder:2019jwn,Dorigoni:2021guq,Collier:2022emf}. Equation \eqref{eq:large_p=N_g0} is a statement about large $N$ factorisation: the (integrated) correlator of maximal-trace operators is simply proportional to the $p=2$ integrated correlator, as expected from the large $N$ statement that $\langle\O_p\O_p\rangle\sim\langle\O_2\O_2\rangle^{p/2}$. The proportionality constant, $p/2$, simply counts the number of $\O_2$ constituents of $\O_p$. An analogous large $N$ factorisation property also holds for extremal two-point functions in $\mathcal{N}=2$ SCFTs at large charge \cite{Beccaria:2018xxl}.

Moving on to the $\mathfrak{g}=1$ term, we have
\begin{align}\label{eq:large_p=N_g1}
\begin{split}
	\mathcal{K}_{\alpha}^{(1)}(\lambda) &=-\frac{\alpha^2}{8}+{1\over 2\pi i} \int_{\Re s =\frac{1}{2}} ds \,{\pi\over\sin(\pi s)} s(1-s)(2s-1)^2\Lambda(1-s)\left(\frac{\lambda}{4\pi}\right)^{s-1}\,k_\alpha^{(1)}(s)\,,
\end{split}
\end{align}
with $k_\alpha^{(1)}(s)$ given in \eqref{eq:k1}. Note that this is no longer proportional to the genus-one $p=2$ term, i.e. an $\mathfrak{g}=1$ analogue of \eqref{eq:large_p=N_g0} does not hold. As usual, both weak and strong coupling expansions are easily obtained from \eqref{eq:large_p=N_g1} by contour deformation of the spectral integral. Here we just quote an alternate exact in $\lambda$ representation in terms of an integral over Bessel functions, which is derived by resummation of the weak coupling expansion and using the identity \eqref{eq:zeta_integral_identity}:
\begin{align}
\begin{split}
	\mathcal{K}_\alpha^{(1)}(\lambda) &= \frac{\alpha^2}{4\pi^2}\int_0^\infty dw~ \frac{w}{\sinh^2(w)}\bigg(w^2\lambda J_0^2\Big(\tfrac{w\sqrt{\lambda}}{\pi}\Big)+(2\pi^2-w^2\lambda)J_1^2\Big(\tfrac{w\sqrt{\lambda}}{\pi}\Big)\\
	&\qquad\qquad\qquad\qquad\qquad\qquad\qquad\qquad-3\pi w\sqrt{\lambda} J_0\Big(\tfrac{w\sqrt{\lambda}}{\pi}\Big)J_1\Big(\tfrac{w\sqrt{\lambda}}{\pi}\Big)\bigg).
\end{split}
\end{align}

Lastly, we comment on the scale of non-perturbative corrections to the strong coupling expansion. Since the large $N$ overlaps $k_\alpha^{(g)}(s)$ are simply related to $g_2^{(0)}(s)$ by a polynomial in $s$, the weak coupling radii of convergence remain unmodified and equal the canonical one, $\pi^2$. Therefore, the non-perturbative scales of exponentially small corrections are always given by $e^{-2\sqrt{\lambda}}$. These statements hold for any genus $g$.

\sec{Comparison to $\cN=2$ SQCD extremal correlators}\label{sec:comparison}

It is natural to compare our results to analogous quantities in $\cN=2$ SCFTs. In this section we compare our integrated correlators $\cG_p^{(N)}(\t)$ in $\cN=4$ SYM to extremal two-point functions in $\cN=2$ SQCD, which depend non-trivially on $\t$. As explained in the introduction, the integrated four-point functions $\cG_p^{(N)}(\t)$ are perhaps the closest non-trivial $\cN=4$ SYM analog of the extremal two-point functions. While the latter have been explicitly computed only for a narrow set of parameters, we perform comparisons where possible, and make some conjectures on the properties of the extremal two-point functions at large charge. 

We consider $\cN=2$ SQCD with $SU(N)$ gauge group and $N_f = 2N$ fundamental flavors. Define the ``extremal'' two-point functions of chiral and anti-chiral primaries,
\e{}{G_{2n}(\t) := \< O_n(0)\overline{O}_n(\i)\>_{\R^4}\,.}
We henceforth specify the operator to be 
\e{}{O_n = [\Tr(\phi^2)]^n\,,\quad n\in\Z_+\,.}
This is an $n$-fold composite of the weight-two chiral primary, with $\phi$ the adjoint scalar in the $\cN=2$ vector multiplet. These $O_n$ are the direct analog of the $\O_p = [\O_2]^{p/2}$ composites studied in the body of this paper, where $p=2n$. They exist for any rank $N\geq 2$. The correlator $G_{2}(\t)$ is related to the Zamolodchikov metric on the $\cN=2$ conformal manifold by the action of supercharges \cite{Baggio:2014ioa}. 

Let us set up a precise comparison. To facilitate this, we will write our $\cN=4$ integrated correlators as $\cG_{2n}^{(N)}(\t)$, i.e. with quantum number
\e{}{p=2n\,,\quad n\in\Z_+\,.}
Now we define the following quantity built from the $\cN=2$ extremal two-point functions:
\e{fdef}{\cF_{2n}^{(N)}(\t) := \log\({G_{2n}(\t)\o G_{2n}^{\cN=4}(\t)}\).}
The denominator in \eqr{fdef}, the ``trivial'' $\cN=4$ SYM two-point function, is a standard normalization factor (see e.g. eq. (4.1) of \cite{Beccaria:2018xxl}). It is this logarithmic quantity that we will compare to our integrated correlators: that is, we compare
\vskip .05 in
\e{}{\text{ \Big($\cF_{2n}^{(N)}(\t)$ in $\cN=2$ SQCD\Big) ~~vs. ~~\Big($\cG_{2n}^{(N)}(\t)$ in $\cN=4$ SYM\Big)}\,.}
\vskip .05 in
\noindent As we will see in what follows, it is the logarithm defined by $\cF_{2n}^{(N)}(\t)$, not the ratio itself, that shares many qualitative features with $\cG_{2n}^{(N)}(\t)$. Note that we take $\t = {\theta\o 2\pi} + {4\pi i\o g^2}$ on both sides.

Before proceeding, we briefly summarize in what regimes $\cF_{2n}^{(N)}(\t)$ has been computed.\foot{$G_{2n}(\t)$ admits a localization expression in general. By ``computed,'' we mean a result that transcends the localization result to make some feature manifest: either written in a closed form, manipulated into a physically insightful form (e.g. a matrix model), reduced from an ($N-1$)-dimensional integral to something simpler, expanded effectively in a particular regime, etc.} First, there are perturbative results $(g^2 \ll 1$) for all $N$ and $n$ \cite{Baggio:2014ioa,Baggio:2015vxa,Bourget:2018obm,Beccaria:2018xxl,Beccaria:2020azj}. Second, there are results in the double-scaling limit of large $n$ with fixed $g^2n$ for low values of $N$ \cite{Bourget:2018obm,Beccaria:2018xxl,Grassi:2019txd,Beccaria:2020azj}. Finally, there are results for large $n$ with fixed $\t$ for $N=2$ \cite{Hellerman:2020sqj,Hellerman:2021yqz,Hellerman:2021duh}. There are no non-perturbative results for $n$ scaling as a power of $N$. 

We now compare these quantities in the three regimes previously considered.

\ssec{Large $p$, finite $N$}\label{sec6.1}

Let us compare $\cF_{2n}^{(N)}(\t)$ and $\cG_{2n}^{(N)}(\t)$ in the double-scaling limit,
\e{dscal}{n\gg 1~,~~ \l_n:=g^2n ~\text{fixed}\,.}
This limit of $\cG_{2n}^{(N)}(\t)$ was considered in Section \ref{sec:p>N}, with the exact leading order result given in \eqr{eq:large_p_g0} for general $N$, and in \eqr{eq:large_p_g0_n23} specialized to $N=2,3$.

The first observation is that both $\cF_{2n}^{(N)}(\t)$ and $\cG_{2n}^{(N)}(\t)$ admit such a limit, and appear to proceed in the same powers of $1/n$. The double-scaling limit of $\cG_{2n}^{(N)}(\t)$ was derived in \eqr{LargeChargeGenusExp}. Based on matrix model analysis in the $N=2$ theory \cite{Grassi:2019txd}, and perturbative analysis for general $N$ \cite{Bourget:2018obm,Beccaria:2018xxl}, it appears that for all $N$,
\e{Fexp}{\cF_{2n}^{(N)}(\t) = \sum_{\mathfrak{g}=0}^\i n^{-\mathfrak{g}}\,\cF_{\mathfrak{g}}^{(N)}(\l_n) + (\text{non-perturbative in $n$})\,.}
The perturbative part matches \eqr{LargeChargeGenusExp}. We will return to the non-perturbative terms below. 

Let us now compare the $\mathfrak{g}=0$ result at weak coupling for all $N$. The expansion of $\cG_{\mathfrak{g}=0}^{(N)}(\l_n)$ was derived in \eqr{eq:large_p_g0_weak}. From eqs. (4.8)--(4.9) of \cite{Beccaria:2018xxl} (see also \cite{Bourget:2018obm,Grassi:2019txd}), 
\es{}{\cF_{\mathfrak{g}=0}^{(N)}(\l_n \ll1) \approx &-18\z(3)\({\l_n\o 8\pi^2}\)^2 + {200(2N^2-1)\o N(N^2+3)} \z(5) \({\l_n\o 8\pi^2}\)^3 \\&- 1225{(8N^6+4N^4-3N^2+3)\o N^2(N^2+1)(N^2+3)(N^2+5)}\z(7) \({\l_n\o 8\pi^2}\)^4 + \ldots.}
At $O(\l_n^{6})$, products of Riemann zeta functions start to appear for all $N>2$, with arbitrarily high degree products appearing in the expansion and violations of uniform transcendentality; this is in contrast to $\cG_{\mathfrak{g}=0}^{(N)}(\l_n)$, whose expansion coefficients at $\l_n \ll 1$ are {\it linear} in zeta functions for all $N$, and uniformly transcendental. We notice that $\cF_{\mathfrak{g}=0}^{(N)}(\l_n \ll1) \sim \l_n^2$, whereas $\cG_{\mathfrak{g}=0}^{(N)}(\l_n \ll1) \sim \l_n$. 

For $N=2$, we can compare exact results for $\cG_{\mathfrak{g}=0}^{(2)}(\l_n)$ and $\cF_{\mathfrak{g}=0}^{(2)}(\l_n)$, using the matrix model calculation of \cite{Grassi:2019txd} for the latter. We have provided three equivalent expressions for $\cG_{\mathfrak{g}=0}^{(2)}(\l_n)$, namely in \eqref{eq:large_p_g0_n23}, \eqref{eq:large_p_g0_N=2} and \eqref{eq:strong_coupling_expansion_N2}. From \cite{Grassi:2019txd}, reproducing their result in our notation,\foot{In Section 4 of \cite{Grassi:2019txd}, the quantity $\cF_{\mathfrak{g}=0}^{(2)}(\l_n)$ is called $\D C_1(\l)$, and the map between couplings is $\l_{\rm GKT} = {1\o 16\pi^2} \l_n$.}
\es{}{\cF_{\mathfrak{g}=0}^{(2)}(\l_n) = -\l_n\({ \log 2\o \pi^2 }\) + \half \log\l_n - \log(4\pi)-1-{\log 2\o 3}+12\log \g_G + \widehat \cF_{\rm inst}^{(2)}(\l_n)\,,}
where
\e{}{\widehat \cF_{\rm inst}^{(2)}(\l_n) = \sum_{k=0}^\i {8\big(K_0((2k+1)\sqrt{\l_n}) + (2k+1)\sqrt{\l_n}K_1((2k+1)\sqrt{\l_n}) \big)\o \pi^2 (2k+1)^2}\,.}
This is functionally similar to the representation \eqref{eq:strong_coupling_expansion_N2}--\eqref{n2polylog} in particular, albeit with some interesting differences at strong coupling:

\begin{itemize}

\item At weak coupling, $\cF_{\mathfrak{g}=0}^{(2)}(\l_n)$ has a convergent expansion with radius $|\l_n| \leq \pi^2$. This is the same radius of convergence obeyed by $\cG_{\mathfrak{g}=0}^{(2)}(\l_n)$, derived in \eqr{eq:radius}. 

\item At strong coupling, $\cF_{\mathfrak{g}=0}^{(2)}(\l_n)$ has a linear term that $\cG_{\mathfrak{g}=0}^{(2)}(\l_n)$ does not. Moreover, though they both admit truncated perturbative expansions at $\l_n\gg1$, their non-perturbative scales differ: 
\es{}{\cF_{\mathfrak{g}=0}^{(2)}(\l_n): \quad &e^{-\sqrt{\lambda_n}}\,,\\
\cG_{\mathfrak{g}=0}^{(2)}(\l_n): \quad &e^{-2\sqrt{\lambda_n}}\,.}
This can be attributed to the difference between fundamental (SQCD) and adjoint (SYM) matter.\footnote{This can be seen from the BPS mass formula \eqref{massformula} (which sets the scale of such corrections) where the electric charges are (half-)integers for matter in the (fundamental) adjoint representation of $SU(2)$.}

\end{itemize}

In the double-scaling limit, $\cF^{(N)}_{2n}(\t)$ will have non-perturbative terms in $n$, represented in \eqr{Fexp}. Based on the likeness between $\cF^{(N)}_{2n}(\t)$ and $\cG^{(N)}_{2n}(\t)$, we expect that these terms are exponentially small in $n$ for all $N$: in particular, 
\e{}{\cF_{2n}^{(N)}(\t) = \sum_{\mathfrak{g}=0}^\i n^{-\mathfrak{g}}\,\cF_{\mathfrak{g}}^{(N)}(\l_n) + O\(e^{-n f(\l_n)}\)\,,}
where $f(\l_n)$ is a function of fixed $\l_n$. This should follow similar logic as recalled around \eqr{npscaleS}, where terms with the above scaling may be seen as other representatives of the ``$SL(2,\Z)$ family'' of instantons. Terms of this sort were computed for $SU(2)$ SQCD in \c{Hellerman:2021duh}.

In Appendix \ref{app:large_p_g1}, we provided the explicit result for $\cG_{\mathfrak{g}}^{(N)}(\l_n)$ at subleading order in $1/n$, i.e. the $\mathfrak{g}=1$ result. For $N=2$, this takes an equivalent form \eqref{eq:large_p_g1_N=2}. These pose a challenge for $\cN=2$ extremal correlators: what are the analogous results for $\cF_{\mathfrak{g}=1}^{(N)}(\l_n)$?

\ssec{Large $p = \a N^2$}

This is the ``gravity regime'', perhaps the most physically interesting one. As noted earlier, $\cF_{2n}^{(N)}(\t)$ has not been studied in this regime. This inspires some new observations. 

Recall that $\cG_{2n}^{(N)}(\t)$ admits a genus expansion \eqref{eq:triple_scaled_expansion} in the triple-scaling limit of large charge, large $N$ and small $g^2$:
\e{triple}{n,N\rar\i~,\quad \a := {2n\o N^2}~ \text{fixed}~,\quad \l := g^2 N ~\text{fixed}\,.}
The leading order result was given in \eqref{eq:genus0_p=N^2}, with $h_\alpha^{(0)}(s)$ recorded in \eqref{eq:h0}.

We claim that $\cF_{2n}^{(N)}(\t)$ also admits a genus expansion in the same limit: that is,
\e{Ftriple}{{\cF_{\a N^2}^{(N)}(\l) = \sum_{\mathfrak{g}=0}^\i N^{2-2\mathfrak{g}} \cF_\a^{(\mathfrak{g})}(\l) + (\text{non-perturbative in $N$})\,.}}
This should be a property of the ``triple-scaled matrix model'' for the extremal correlators in the limit \eqr{triple}. 

The expansion \eqr{Ftriple} is supported by available evidence in the literature. The weak coupling expansion for finite $n$ and $N$ through $O(g^{10})$ was given in \cite{Beccaria:2018xxl}, eqs. (2.16), (4.6) and (4.7) (building upon \cite{Bourget:2018obm}). One readily confirms with that data that \eqr{Ftriple} holds to arbitrary order in $\mathfrak{g}$, at least through $O(\l^5)$. At leading order, for example, 
\es{Fgenuspert}{\cF_\a^{(0)}(\l\ll1) = &-\frac{9 \alpha  (\alpha +1) \lambda ^2 \z(3)}{128 \pi ^4}+\frac{5 \alpha  \left(10 \alpha ^2+15 \alpha +6\right) \lambda ^3 \z(5)}{512 \pi ^6}\\&+\frac{9 \left(18 \alpha ^3+27 \alpha ^2+11 \alpha \right) \lambda ^4 \z(3)^2}{16384 \pi ^8}-\frac{135 (\alpha +1) \left(10 \alpha ^3+10 \alpha ^2+3 \alpha \right) \lambda ^5 \z(3) \z(5)}{32768 \pi ^{10}}\\
&+O(\l^6)\,,}
with similar expressions at $\mathfrak{g}>0$. 

This organization of $\cF_{2n}^{(N)}(\t)$ into a genus expansion seems to have gone unnoticted. It holds only for the logarithm, involving non-trivial cancellations among different orders of $G_{2n}(\t)/ G_{2n}^{\cN=4}(\t)$; this is intuitive if we think of $\cF_{2n}^{(N)}(\t)$ as a free energy with extra operator insertions. See e.g. \cite{Beccaria:2022kxy} for recent work in a somewhat related arena.

It is an attractive but difficult problem to derive $\cF_\a^{(0)}(\l)$ at finite $\l$ in $\cN=2$ SQCD, in analogy to the results for $\cG_\a^{(0)}(\l)$ in \eqref{eq:genus0_p=N^2} and \eqref{besselrepfinal}. Short of this, one might aim to find the analog of the non-perturbative scale \eqref{eq:scale_new} involving the large charge dressing factor $R_\a$. The planar strong coupling limit of $\cN=2$ SQCD is {\it not} dual to supergravity \cite{Gadde:2009dj} (for example, $a \neq c$ at large $N$), so a bulk interpretation of the dressing factor $R_\a$ would necessarily be in a stringy regime.

If we do {\it not} scale $g^2$ to zero, instead keeping both $\a$ and $\t$ fixed as $n,N \rar\i$, what do we expect of $\cF_{2n}^{(N)}(\t)$? It seems very likely that it will be more complicated than $\cG_{2n}^{(N)}(\t)$, whose exact leading order expression was given in Section \ref{sec:p~N2}. For one, the expansion coefficients in \eqr{Fgenuspert} imply that in the spectral decomposition, the Eisenstein overlap will itself generate Riemann zeta functions when evaluated on the integer residues. We are also not aware of an argument that the Maass cusp form overlap would be nonzero for $\cF_{2n}^{(N)}(\t)$, unlike for $\cG_{2n}^{(N)}(\t)$.

\ssec{Large $p = \a N$} 

As noted earlier, $\cF_{2n}^{(N)}(\t)$ has not been studied in this regime. 

We recall from Section \ref{sec:p~N} that this limit of $\cG_{2n}^{(N)}(\t)$ is rather simple. At leading order, it is just a constant times the planar $n=1$ result (see \eqr{eq:k0}). The intuition for this was that since $n \ll N^2$, the leading-order result for $n \sim N \gg 1$ is just the $n \gg1$ limit of the planar result, and the latter obeys a large $N$ factorization. In the 't Hooft limit of fixed $\l = g^2 N$, this implied \eqref{eq:large_p=N_g0}, which we reproduce here:
\e{}{\cG_{\a N}^{(N)}(\l) \sim {N \a\o2} \,\cG_2^{(\mathfrak{g}=0)}(\l) + O(N^0)\,.}

We claim that this same structure holds for $\cF_{2n}^{(N)}(\t)$, both for its spectral overlaps and in the 't Hooft limit. The latter implies that
\e{}{\cF_{\a N}^{(N)}(\l) \sim  {N \a\o2} \cF_2^{(\mathfrak{g}=0)}(\l) + O(N^0)\,.}
In other words, when the charge $n \sim N$, the (logarithm of the) extremal correlator is simply a constant rescaling of the (logarithm of the) Zamolodchikov metric. This holds for any $1 \ll n \ll N^2$. Once again, this is supported by studying the $n={\a\o2} N$ scaling limit of perturbative computations. From Section 4 of \cite{Beccaria:2018xxl}, one readily confirms the above identity. The logarithm is crucial for this correspondence. At higher orders in $1/N$, $\cG_{\a N}^{(N)}(\l)$ again has a simple relation to $\cG_2^{(\mathfrak{g}=0)}(\l)$, as seen in \eqr{eq:k1}; we expect that $\cF_{\a N}^{(N)}(\l)$ should have a similar property. 

\section{Exact solution for odd-$p$ maximal-trace correlators}\label{app:p3}
In this final section we present a generalisation of our results to a \textit{second infinite tower} of integrated correlators, where instead of charged operators \eqr{opdef} we now consider\footnote{Extremal two-point functions involving such operators have also been studied in $\cN=2$ SQCD \cite{Beccaria:2018xxl,Beccaria:2020azj}}
\begin{align}\label{opdef2}
\widetilde\O_p := \O_3\big[\O_2\big]^{\frac{p-3}{2}}\,,\qquad p-1\in2\Z_+\,.
\end{align}
This defines the \textit{odd-$p$ family} of integrated maximal-trace correlators. Their structure turns out to be very similar to the even-$p$ family $\cG_p^{(N)}(\t)$ considered in previous sections, and in fact the analysis proceeds in an identical manner.

Integrated correlators \`a la \eqref{eq:integrated_correlator} involving such operators may again be computed from supersymmetric localization via the prescription in \cite{Paul:2022piq,Binder:2019jwn,Gerchkovitz:2016gxx}. We find that they admit the spectral integral representation \eqref{eq:spectral_decomp} where the Eisenstein overlaps\footnote{An explicit one-instanton calculation provides evidence that the overlap with Maass cusp forms vanishes \cite{Paul:2022piq}.}, which we denote by $\widetilde g_p^{(N)}(s)$, satisfy a second order recursion relation of the same form as \eqref{eq:recursion_maxtrace}:
\begin{align}\label{eq:recursion_maxtrace3}
s(1-s)\widetilde g_{p-2}^{(N)}(s) = -\widetilde \kappa_p \(\tilde g_{p}^{(N)}(s)-\widetilde g_{p-2}^{(N)}(s)\) + \widetilde \kappa_{p-2} \(\widetilde g_{p-2}^{(N)}(s)-\widetilde g_{p-4}^{(N)}(s)\)+\frac12(N^2-1)g_2^{(N)}\,,
\end{align}
with the only difference that the coupling $\widetilde \kappa_p$ is now given by
\be
\label{base3coupling}
\widetilde{\kappa}_p=\frac{1}{4}(p-3)\left(N^2+p\right).
\ee
The recursion \eqref{eq:recursion_maxtrace3} starts at $p=5$, with the $p=3$ overlap $\widetilde g_3^{(N)}(s)$ as an initial condition. The solution of \eqref{eq:recursion_maxtrace3} is 
\be\label{tildegpsol}
\widetilde g_{p}^{(N)}(s)=\widetilde F_p(N, s)\, \widetilde g_{3}^{(N)}(s)+\left[1-\widetilde F_p(N, s)\right] \frac{N^2-1}{2 s(1-s)}\, g_2^{(N)}(s)\,,
\ee
where  
\be
\boxed{\widetilde F_p(N,s)= \, _3F_2\left(\frac{3-p}{2},1-s,s;1,\frac{N^2+5}{2};1\right).}
\ee
The $p=3$ overlap $\widetilde g_{3}^{(N)}(s)$ is found to satisfy a first-order recursion relation given in (1.4) of \cite{Paul:2022piq}, which in the notations of this paper reads
\begin{align}
	N^2(N+3)\widetilde g_3^{(N+1)}(s)-(N+1)^2(N-2)\widetilde g_3^{(N)}(s) = 3\left[ N^2 g_2^{(N+1)}(s)+(N+1)^2 g_2^{(N)}(s)\right].
\end{align}
The solution of this recursion is 
\begin{align}
\widetilde g_3^{(N)}(s) = \frac{N}{84(N+1)(N+2)}&\bigg[6 \big(5 N (s^2-s+2)-8 (s-4) (s+3)\big)\,_3F_2(3-N,1-s,s;4,4;1)\no\\
&+14 (N-3) (s^2-s+18) \, _3F_2(4-N,1-s,s;4,4;1)\\
&-(N-3) (s-4)^2 (s+3)^2 \, _3F_2(4-N,1-s,s;5,5;1)\bigg]\,.\no
\end{align}
These results are manifestly symmetric under $s\leftrightarrow 1-s$. Similarly to the solution \eqr{g2Expr} for $g_p^{(N)}(s)$, the solution \eqr{tildegpsol} is analytic in an extended regime which covers the full physical parameter space $p\geq 3$ and $N\geq 2$. 

The analysis of the large charge expansion of this observable is no different from that of the case discussed in previous sections. Comparing the form of the spectral overlaps $g_p^{(N)}(s)$ and $\widetilde g_p^{(N)}(s)$, they are structurally almost identical. In particular, there are again three distinct regimes of large $p$: namely, $p\gg N^2$, $p\sim N^2$ and $p\ll N^2$. 

For example, at large $p$ and finite $N$, the function $\widetilde F_p(N,s)$ organizes into the expansion \eqref{LargePFiniteNOverlap}: the existence of the double-scaled large charge limit of fixed $\l_p$ is again manifest. The leading result at $g=0$ is given by 
\begin{align}\label{LargeOddPFiniteNOverlapg0}
\widetilde F^{(0)}(N, s) =\frac{2^{s-1} \Gamma \left(\frac{N^2+5}{2} \right)  \Gamma \left(s-\frac{1}{2}\right)}{\sqrt{\pi } \Gamma (s) \Gamma \left(\frac{N^2+3}{2}+s\right)}\,.
\end{align}
Again, the strong coupling expansion in the double-scaled regime truncates for even $N$. The non-perturbative corrections are controlled by the same scale. 

At large $p$ and large $N$ (for any relative scaling), $\widetilde F_p(N,s)$ and $F_p(N,s)$ become identical at leading order. Therefore, the leading-order results for the respective integrated correlators are almost identical, with small differences arising only due to the details of \eqr{g2Expr} and \eqr{tildegpsol}. For example, in the $p\sim N^2$ regime, the large charge dressing factor $R_\alpha$ defined in \eqref{eq:Ralpha}, and the screened non-perturbative scale $\L^2_\a(\l)$ defined in \eqref{eq:scale_new}, both carry over to this odd-$p$ case.\foot{The large $N$ expansion of $\widetilde g_3^{(N)}(s)$ was carried out in \cite{Paul:2022piq}, see eqs. (5.1)-(5.3) therein. In particular, the large $N$ expansion of $\widetilde g_3^{(N)}(s)$ is proportional to that of $g_2^{(N)}(s)$, which implies that the asymptotic behaviour of odd-$p$ spectral overlaps (in the spectral parameter $s$) is identical to the even-$p$ case.} 

These similarities are intuitively expected: for the even- and odd-$p$ maximal-trace families of correlation functions, the large charge limit is obtained by dialing up the number of $\O_2$ constituents, and small changes of the ``base-operator'' used to form the charge $p$ operator (given by $\O_p$ or $\widetilde{\O}_p$, respectively) should not affect the large charge asymptotics. This uniformity is also seen in the coupled harmonic oscillator representation, where the couplings $\kappa_p$ in \eqref{eq:beta} and $\widetilde \kappa_p$ in \eqref{base3coupling} become identical at large $p$.

In fact, one may form a more general operator $\O_p^{(m)}$ as a composite of $(p-m)/2$ copies of $\O_2$ and a base-operator $\O_m$ of finite dimension $m$. As shown in \cite{Gerchkovitz:2016gxx}, for any integer $m$ there exists a suitable definition of $\O_m$ such that the semi-infinite towers $\O_p^{(m)}$ mutually decouple. These considerations lead us to {\it conjecture} that the $\<\O_2\O_2\O_p^{(m)}\O_p^{(m)}\>$ integrated correlators are governed by a coupled harmonic system analogous to \eqref{eq:lattice_chain} -- equivalently, a recursion relation \eqref{eq:recursion_maxtrace} -- with dependence on the choice of base-operator entering only through some generalised nearest-neighbour couplings $\kappa_p^{(m)}$.\footnote{We note that an even stronger uniformity is known to hold for extremal two-point functions in $\mathcal{N}=4$ SYM, where the Toda chain is structurally \textit{independent} of the choice of the base-operator (see (3.37) in \cite{Gerchkovitz:2016gxx}).} Moreover, we expect that to leading order in the large charge limit, these couplings become $m$-independent: that is,
\be
\kappa^{(m)}_{p\gg 1} = \frac{p^2}{4}~,\qquad \kappa^{(m)}_{N^2\gg p\gg 1} =\frac{pN^2}{4}\,.
\ee
This expectation is based on the physical argument given above that the base-operator does not affect the large charge asymptotics -- which is indeed confirmed by the two explicit examples presented in this paper. The same reasoning further suggests that also in the gravity regime $p\sim N^2$, the leading-order term is universal. A first principles derivation of the coupled harmonic oscillator description of $\cN=4$ SYM integrated correlators would hopefully shed light on these expectations.

\section*{Acknowledgements}

We thank Scott Collier, Abhijit Gadde, Francesco Galvagno, Simeon Hellerman, Dileep Jatkar, Shota Komatsu, Grisha Korchemsky, Gautam Mandal, Mark Mezei and Ioannis Tsiares for discussions. This research was supported by ERC Starting Grant 853507, and in part by the National Science Foundation under Grant No. NSF PHY-1748958.

\appendix
\section{Properties of Eisenstein series}\label{app:Eisensteins}

The real-analytic (non-holomorphic) Eisenstein series is the unique modular-invariant solution to the following eigenvalue equation

\e{EinDef}{\Delta_\tau E_s(\tau)=s(1-s) E_s(\tau)\,,}
where $\Delta_\tau$ is the scalar hyperbolic Laplacian
\e{}{\Delta_\tau=-y^2\left(\partial_x^2+\partial_y^2\right),}
where $\tau=x+iy$. The completed Eisenstein series is given by
\e{EinCompleted}{E_s^*(\tau):=\Lambda(s) E_s(\tau)\,,}
where $\Lambda(s)$ is the completed Riemann zeta function
\e{ZetaCompleted}{\Lambda(s):=\pi^{-s} \Gamma(s) \zeta(2 s)=\Lambda\left(\frac{1}{2}-s\right).}
The completed Eisenstein series satisfies the functional equation
\e{EinReflection}{E_s^*(\tau)=E_{1-s}^*(\tau)\,,}
which is manifest from its Fourier expansion
\e{EinDecomp}{E_s^*(\tau)=\Lambda(s) y^s+\Lambda(1-s) y^{1-s}+\sum_{k=1}^{\infty} 4 \cos (2 \pi k x) \frac{\sigma_{2 s-1}(k)}{k^{s-\frac{1}{2}}} \sqrt{y} K_{s-\frac{1}{2}}(2 \pi k y)~.}
There is a simple pole at $s=1$ with constant residue,
\e{Eispole}{\Res_{s=1}E_s^*(\t) = {1\o2}}
The Eisenstein series also admits a representation in terms of the following lattice sum 
\e{EinLatt}{E_s^*(\tau) = \frac{\Gamma(s)}{2} \sum_{(m,n)\neq (0,0)}  Y_{m n}(\tau)^{-s}~,\qquad Y_{m n}(\tau) = \pi \frac{|m+n\tau|^2}{y}~.}
We refer to Section 2 of \cite{Collier:2022emf} for further details.

\section{Various expansions of Eisenstein overlaps}\label{app:Fp_expansions}

In this appendix we show how to expand the overlaps given in \eqref{g2Expr} for various regimes. We analyse $F_p(N,s)$ and $g_2^{(N)}(s)$ separately. 

\subsection{Expansions of $F_p(N,s)$}

The first step is to express the generalized hypergeometric function in \eqref{eq:Fp_closed} as the following double integral

\begin{align}\label{DoubleIntRep}
\,_3F_2\(-\frac{p}{2},s,1-s;1,\frac{N^2-1}{2};1\) &=\frac{\Gamma \left(\frac{N^2-1}{2}\right)}{\Gamma^2 (1-s) \Gamma (s) \Gamma \left(\frac{N^2-3}{2}+s\right)} \x\no \\
& \int_0^1d\zeta_1\int_0^1  d\zeta_2~ \left(1-\zeta _1 \zeta _2\right){}^{\frac p2} \zeta _1^{-s} \left(1-\zeta _2\right){}^{-s} \zeta _2^{s-1} \left(1-\zeta _1\right){}^{\frac{1}{2} \left(N^2-5\right)+s}~.
\end{align}
We now analyze this in various regimes of large $N$ and $p$.

\sssec{$p \gg N^2$}
First we look at large $p$ with $N$ fixed, which was analyzed in Section \ref{sec:p>N}. Notice that the $p$ dependence is only through the $ (1-\z_1\z_2)^{\frac p2}$ term. At large $p$, such a term is exponentially suppressed at a generic point in the integration domain. Dominant contributions come only from the region where $\zeta_1\zeta_2$ is of the order $1/p$. This splits the integration domain into two parts: one where  $\zeta_1\sim 1/p$ with $\z_2$ finite and the other where $\zeta_2\sim 1/p$ with $\z_1$ finite. In the former case we proceed by rescaling $\zeta_1$ as
\be
\zeta_1 = \frac{x}{p}~,
\ee
and expanding the integrand in \eqref{DoubleIntRep} at large $p$. This gives the following leading order result 
\be
\int_0^\infty dx \int_0^1 d\zeta_2 ~ \left(1-\zeta _2\right){}^{-s} \zeta _2^{s-1} x^{-s} e^{-\frac{1}{2} \zeta _2 x} = \frac{2^{s-1} \Gamma^2(1-s) \Gamma \left(s-\frac{1}{2}\right)}{\sqrt{\pi }}\,.
\ee
Combining this with the $p$-independent prefactor appearing in the first line of \eqref{DoubleIntRep} we obtain the genus-zero result mentioned in \eqref{LargePFiniteNOverlapg0} in the main text:
\be
F^{(0)}(N,s)=\frac{2^{s-1}\Gamma\left(\frac{N^2+1}{2}\right) \Gamma\left(s-\frac{1}{2}\right)}{\sqrt{\pi}(s-1) \Gamma(s+1) \Gamma\left(\frac{N^2-3}{2}+s\right)}\,.
\ee
Repeating the analysis for the $\zeta_2\sim1/p$ edge we find the $s\leftrightarrow 1-s$ reflected partner of the above result. In this way one finds that the large $p$ expansion of $F_p(N,s)$ organizes into the genus expansion shown in \eqref{LargePFiniteNOverlap}, where results for higher genera were recorded in \eqr{eq:Fp_higher_g}. An alternative way to generate higher genus overlaps to is take the seed expression \eqref{LargePFiniteNOverlapg0} and use the recursion formula \eqref{eq:recursion_maxtrace}. 

\sssec{$p \ll N^2$}
Next we look at the case where $1 \ll p \ll N^2$, with $p = \alpha N$ being a special case studied in Section \ref{sec:p~N}. Here we have to focus on the contribution from the factor $\left(1-\zeta _1\right){}^{\frac{1}{2} \left(N^2-5\right)+s}$ which is exponentially suppressed at large $N$ unless $\z_1\sim 1/N^2$. Therefore, we localize the integration domain to the edge by again defining
\be
\zeta_1 = \frac{x}{N^2}~.
\ee
Note that this takes care of the $ (1- \z_1\z_2)^{\frac p2}$ term which automatically gives a finite contribution when $p \ll N^2$. Expanding the integrand at large $N$ and integrating back term-by-term we obtain the following large $N$ expansion for $F_p(N,s)$
\begin{align}\label{F_LargeN}
\begin{split}
F_p(N, s)= &~\frac{p}{2}+\frac{p(p-2)}{8}(s+1)(s-2) N^{-2} \\
& +\frac{p(p-2)}{72}(s+1)(s-2)[(p-4) s(s-1)-3(2 p-5)] N^{-4} \\
& +\frac{p(p-2)}{1152}(s+1)(s-2)\left[(p-4)(p-6) s^3(s-2)-(p-4)(17 p-38) s^2\right. \\
& \left.\quad+2(p-4)(9 p-22) s+24\left(3 p^2-14 p+14\right)\right] N^{-6} +O\left(N^{-8}\right).
\end{split}
\end{align}
This reproduces eq. (6.20) in \cite{Paul:2022piq}. The subsequent large $p$ expansion trivially follows as long as $p \ll N^2$ (at a given order in $1/N$, the large $p$ expansion terminates). In the special case of $p=\alpha N$, 
the result is
\begin{align}\label{eq:F_large_p=N}
F_{\alpha N}(N,s) =&~\frac{\alpha }{2} N + \frac{1}{8} \alpha ^2 (s-2) (s+1) \\
&+\frac{1}{72} \alpha  (s-2) (s+1) \left(\alpha ^2 s^2-6 \alpha ^2-\alpha ^2 s-18\right)\frac{1}{N}\no \\
&+\frac{\alpha ^2 (s-2) (s+1)}{1152} \left(72 \alpha ^2+\alpha ^2 s^4-2 \alpha ^2 s^3-17 \alpha ^2 s^2-96 s^2+18 \alpha ^2 s+96 s+432\right)\frac{1}{N^2}\no\\
&+O(N^{-3})~.\no
\end{align}

\sssec{$p \sim N^2$}

We now come to the especially interesting regime $p= \a N^2 \gg 1$ with fixed $\a$, which was analyzed in Section \ref{sec:p~N2}. In this case, the two terms $ (1-\z_1 \z_2)^{\frac p2}$ and $\left(1-\zeta _1\right){}^{\frac{1}{2} \left(N^2-5\right)+s}$ compete. The integrand in \eqref{DoubleIntRep} is suppressed everywhere except for $\z_1\sim 1/N^2$. Repeating the analysis we find the following expansion
\begin{align}
\label{eq:c_alpha0}
\begin{split}
F_{\alpha N^2}(N,s) = \sum_{m=0}^{\infty} c_\alpha^{(m)}(s) N^{2-2 m} \,,
\end{split}
\end{align}
with the first few coefficient functions $c_\alpha^{(m)}(s)$ given by
\begin{align}\label{eq:c_alpha}
c_\alpha^{(0)}(s) &=\frac{1-\,_2F_1(1-s,s;1;-\alpha )}{2 s(1-s)}\,,\no\\
c_\alpha^{(1)}(s) &= \frac{1}{4(\alpha+1) s(1-s)}\left[\alpha(s-2)(s+1)\left(s^2-s+1\right){ }_2 F_1(1-s, s ; 3 ;-\alpha)\right. \no \\
&\quad \left.-2(\alpha(s-2)(s+1)-1){ }_2 F_1(1-s, s ; 2 ;-\alpha)-2(\alpha+1)\right], \\
c_\alpha^{(2)}(s) &= \frac{\alpha(s-2)(s+1)}{288(\alpha+1)^3}\left[3 \alpha(s-4)(s+3)((\alpha+1)(s-1) s-5 \alpha-4){ }_2 F_1(1-s, s ; 5 ;-\alpha)\right. \no \\
&\quad \left.+4\left(2 \alpha^2\left(5 s^2-5 s-21\right)+9 \alpha\left(2 s^2-2 s-7\right)+2(2 s-5)(2 s+3)\right){ }_2 F_1(1-s, s ; 4 ;-\alpha)\right].\no
\end{align}

The $\alpha\to0$ limit of these results agrees with the leading-order result in the $p \ll N^2$ regime. (This is seen in the expression \eqref{smallalphaexpansion}.) Likewise, the $\alpha\to\infty$ limit agrees with the leading-order result in the $p \gg N^2$ regime.\footnote{The easiest way to see this is to expand the following representation in large $\alpha$:
\be
{ }_2 F_1(1-s, s ; 1 ;-\alpha) = \frac{1}{\pi}\int_0^\pi dt \left(2 \alpha -\sqrt{(2 \alpha +1)^2-1} \cos (t)+1\right)^{s-1}~.
\ee
}

We note that unlike the $p\gg N^2$ regime, the results for higher order terms in \eqref{F_LargeN}, \eqref{eq:F_large_p=N} and \eqref{eq:c_alpha0} cannot be obtained directly from the recursion formula \eqref{eq:recursion_maxtrace}.

\subsection{Large $N$ expansion of $g_2^{(N)}(s)$}\label{app:p=2_recap}

The large $N$ expansion of the $p=2$ overlap $g_2^{(N)}(s)$ can be done in exactly the same manner. Its integral representation is given by
\begin{align}\label{eq:g2_closed_1}
	g_2^{(N)}(s) &= \frac{N}{N+1}\,_3F_2\(2-N, s,1-s;3,2;1\) \\&= \frac{2 N \sin ^2(\pi  s)}{\pi ^2 (N+1) (s-2) (s-1) s} \int_0^{1} d\z_1 \int_0^{1} d\z_2 \left(1-\zeta _1 \zeta _2\right){}^{N-2} \left(1-\zeta _1\right){}^s \zeta _1^{-s} \left(1-\zeta _2\right){}^{2-s} \zeta _2^{s-1}\,~.
\end{align}
At large $N$, the integral is dominated by the regions where $\z_1\sim 1/N$ with $\z_2$ finite and $\z_2\sim 1/N$ with $\z_1$ finite. This is analogous to the large $p$ expansion discussed in the previous section. Proceeding along similar lines one finds
\begin{align}\label{eq:g2_large_N_expansion}
	g_2^{(N)}(s) = \sum_{g=0}^{\infty} N^{-2g}\Big[N^{s-1}g_2^{(g)}(s) + N^{-s}g_2^{(g)}(1-s)\Big],
\end{align} 
where 
\begin{align}
\label{eq:g2_higher-genus}
\begin{split}
g_2^{(0)}(s) &= \frac{2^{2 s-1} \Gamma \left(s-\frac{1}{2}\right)}{\sqrt{\pi } \Gamma (s+1) \Gamma (s+2)}\,,\\
g_2^{(1)}(s) &= \frac{(s-4)_3 (s+3)}{24 (2 s-3)}\,g_2^{(0)}(s)\,,\\
g_2^{(2)}(s) &= \frac{(s-6)_5 (5 s^3+8 s^2-37 s-120)}{5760 (2 s-5) (2 s-3)}\,g_2^{(0)}(s)\,,\\
g_2^{(3)}(s) &=\frac{(s-8)_7 (35 s^5-77 s^4-401 s^3-643 s^2+2166 s+7560)}{2903040 (2 s-7) (2 s-5) (2 s-3)}\,g_2^{(0)}(s)\,.
\end{split}
\end{align}
The higher order terms can be alternatively computed from the recursion \eqref{eq:recursion_g2} given the seed $g_2^{(0)}(s)$.

\subsection{Assembling the large $N$ expansion of $g_p^{(N)}(s)$ in the $p=\alpha N^2$ regime}\label{app:assembly}
In the $p=\alpha N^2$ double-scaling regime, the maximal-trace overlaps $g_p^{(N)}(s)$ admit a large $N$ genus expansion of the form
\begin{align}\label{eq:large_p=N^2_v2}
	g_{\alpha N^2}^{(N)}(s)=\sum_{g=0}^\infty N^{2-2g}\big[N^{s-1}h_\alpha^{(g)}(s)+N^{-s}h_\alpha^{(g)}(1-s)\big]\,,
\end{align}
which was already presented in \eqref{eq:large_p=N^2}. The genus-$g$ overlaps $h_\alpha^{(g)}(s)$ are easily obtained by plugging in the large $p=\alpha N^2$ expansion of $F_p(N,s)$ from \eqref{eq:c_alpha0} and the usual large $N$ expansion of $g_2^{(N)}(s)$ from \eqref{eq:g2_large_N_expansion} into the defining relation
\begin{align}
	g_p^{(N)}(s) = F_p(N,s)\,g_2^{(N)}(s)\,.
\end{align}
Re-expanding the product and comparing to \eqref{eq:large_p=N^2_v2} determines $h_\alpha^{(g)}(s)$ in terms of the $c_\alpha^{(m)}(s)$ and $g_2^{(g)}(s)$ derived above:
\begin{align}
\boxed{h_\alpha^{(g)}(s) = \sum _{m=0}^g c_\alpha^{(m)}(s)\,g_2^{(g-m)}(s)}
\end{align}
For completeness, let us explicitly spell out the first few orders:
\begin{align}\label{higherggrav}
\begin{split}
	h_\alpha^{(0)}(s) &= c_\alpha^{(0)}(s)\,g_2^{(0)}(s)\,,\\
	h_\alpha^{(1)}(s) &= c_\alpha^{(1)}(s)\,g_2^{(0)}(s)+c_\alpha^{(0)}(s)\,g_2^{(1)}(s)\,,\\
	h_\alpha^{(2)}(s) &= c_\alpha^{(2)}(s)\,g_2^{(0)}(s)+c_\alpha^{(1)}(s)\,g_2^{(1)}(s)+c_\alpha^{(0)}(s)\,g_2^{(2)}(s)\,,
\end{split}
\end{align}
where the necessary data for $c_\alpha^{(m)}(s)$ and $g_2^{(g)}(s)$ is given in equations \eqref{eq:c_alpha} and \eqref{eq:g2_higher-genus}, respectively.

\section{Some details about $H^{(N)}(\tau)$}\label{appendix:H}

The $H^{(N)}(\tau)$ function was defined in \eqref{HFunction}, which we reproduce here:
\be\label{HFunctionapp}
H^{(N)}(\tau) := \frac{1}{4\pi i}\int_{\text{Re}\,s=\frac{1}{2}}ds\,\frac{\pi}{\sin(\pi s)}(2s-1)^2g_2^{(N)}(s)\,E^*_s(\tau)\,.
\ee
As this definition makes clear, $H^{(N)}(\tau)$ is formally equal to the action of the inverse Laplacian on $\cG_2^{(N)}(\t)$ minus its average:
\es{Hshift}{\D_\t H^{(N)}(\t) = \cG_2^{(N)} - \< \cG_2^{(N)} \>\,,}
where in the normalization conventions of this paper,
\e{}{\< \cG_2^{(N)} \> = {N(N-1)\o 4}{2\o N^2-1}\,.}
Note that ${N^2-1\o 2}H^{(N)}(\t)$ is almost the inhomogeneous part of the oscillator $\cQ_p^{(N)}(\t)$ defined in \eqr{Qdef}, differing only by the shift of the average term in \eqr{Hshift}.

We may compute the zero mode of $H^{(N)}(\t)$ by inserting the zero mode of the completed Eisenstein series, $E_{s,0}^*(y)$, given by the first two terms in equation \eqref{EinDecomp}. Inserting this into \eqref{HFunction} thus gives
\begin{align}\label{Hzeromode}
H^{(N)}_0(y) &=\frac{1}{2\pi i}\int_{\text{Re}\,s=\frac{1}{2}}ds\,\frac{\pi}{\sin(\pi s)}(2s-1)^2g_2^{(N)}(s)  \Lambda (1-s)y^{1-s}\,,
\end{align}
where $g_2^{(N)}(s)$ is the generalized hypergeometric function given in \eqr{eq:g2_closed}. 
This contains the perturbative ($y \gg 1$) part of this function. To develop the $y \gg 1$ expansion, we close the contour towards the right. Using the explicit expression \eqr{eq:g2_closed}, and the result 
\e{3F2deriv}{\[{\p\o \p s}\,{}_3F_2(2-N,s,1-s;2,3;1)\]_{s=1} = {2 (N+1) H_N-5 N+1\o 2(N-1)}\,,}
we find the following perturbative expansion of $H^{(N)}(\t)$: 
\es{Hzeromode2}{H^{(N)}_0(y) &= {N\o 4(N^2-1)}\Big(2 (N-1) (\gamma_E-\log (4 \pi  y))+2 (N+1) H_N  +3 N-7\Big) \\
&- \frac{N}{N+1}\sum_{s=2}^\infty (-1)^s (2s-1)^2\,_3F_2\(2-N, s,1-s;3,2;1\) \Lambda(1-s) y^{1-s}\,.}
The logarithm comes from the fact that the integrand of \eqr{Hzeromode} has a double pole at $s=1$. This was used in Section \ref{sec:3.2}, and ultimately leads to the $\log\l_p$ term in the double-scaling limit of large $p$, fixed $\l_p = g^2 p$ for any $N$: isolating the relevant terms of \eqr{genericNvsc},
\es{Hloglp}{{N(N-1)\o 4} \log \({p\o 2}\) + {N^2-1\o 2}H^{(N)}(\t) &\supset {N(N-1)\o 4} \log \(p\o 8\pi y\)\\&\mapsto {N(N-1)\o 4}\log\({\l_p \o 32\pi^2}\),}
where the second line rewrites the first in the double-scaling limit. This produces the logarithm in \eqr{eq:large_p_g0}. 

Let us make another comment on the logarithmic behavior. It is crucially tied to the shift in \eqr{Hshift}. In particular, the definition of $H^{(N)}(\t)$ implies that
\e{delH}{\D_\t H^{(N)}(\t) = \cG_2^{(N)}(\t) - {N(N-1)\o 4}{2\o N^2-1}\,.}
Let us examine this equation near the cusp. Noting that $\cG_2^{(N)}(y\gg1) \sim y^{-1}$, the large $y$ behavior of $\D_\t H^{(N)}(\t)$ is therefore given by the second term on the right-hand side of \eqr{delH}:
\e{}{\D_\t H^{(N)}(\t)\Big|_{y \gg 1} =  -{N(N-1)\o 4}{2\o N^2-1} + \O(y^{-1})}
From \eqr{Hzeromode2}, the leading $y \gg 1$ behavior of $H^{(N)}(\t)$ is given by the $\log y$ term (nonzero modes being exponentially suppressed). Applying $\D_\t = -y^2 (\p_x^2 + \p_y^2)$ to that term yields exactly the result above. 

The series in the second line of \eqr{Hzeromode2} is asymptotic, but Borel summable. Employing the $SL(2,\Z)$ Borel transform \cite{Collier:2022emf} in which we divide the perturbative series \eqr{Hzeromode2} by $\L(1-s)$,
\e{}{B[H^{(N)}_0; \xi] := - \frac{N}{N+1}\sum_{s=2}^\infty (-1)^s (2s-1)^2\,_3F_2\(2-N, s,1-s;3,2;1\)\xi^{s-1}\,,}
allows for the simplest resummation. The sum may be easily evaluated for any fixed $N$, and inverted to give the Borel resummation of the original series -- call it $\widehat{H}^{(N)}_0(y)$, following \cite{Collier:2022emf} -- using
\e{}{\widehat{H}^{(N)}_0(y) = y^{1\o2}\int_0^\i {d\xi\o \xi^{3\o2}}\({\theta_3(y \xi)-1\o 2}\)B[H^{(N)}_0; \xi]\,,}
where $\theta_3$ is the Jacobi theta function. Let us list the $SL(2,\Z)$ Borel transforms for $N=2,3$:
\es{}{B[H^{(2)}_0; \xi] &= \frac{2 \xi  \left(\xi ^2+2 \xi +9\right)}{3 (\xi +1)^3}\\
B[H^{(3)}_0; \xi]&= \frac{3 \xi  \left(\xi ^4+4 \xi ^3+17 \xi ^2+10 \xi +12\right)}{4 (\xi +1)^5}}
%

\section{Exponentially small corrections at large $p$, fixed $\t$ }\label{App3}

\begin{figure}[t]
\begin{center}
	\begin{minipage}[h]{4cm} 
		\centering 
		\includegraphics[scale=0.9]{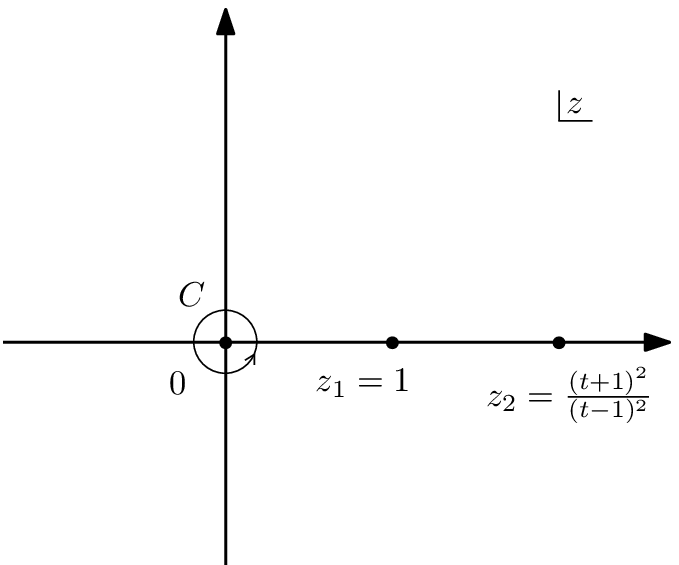}
		\caption*{(a)}
		\label{ContourFig1}
	\end{minipage} 
	\hspace{4cm} 
		\begin{minipage}[h]{7cm} 
		\centering 
		\includegraphics[scale=0.9]{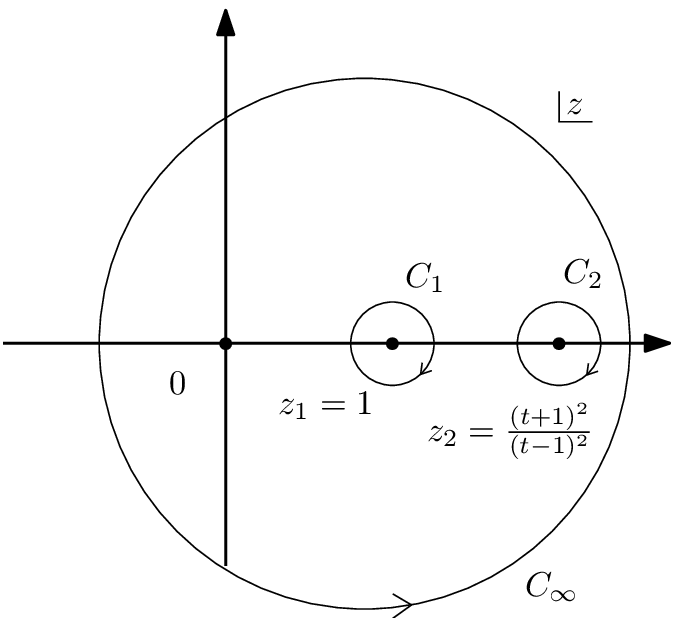}
		\caption*{(b)}
		\label{ContourFig2}
	\end{minipage} 		
\end{center}
\caption{(a) The $B(z;t)$ function appearing in \eqref{N=2GenFun} has poles at $z_1$ and $z_2$. The contour $C$ encloses the $z=0$ pole of the integrand appearing in \eqref{N=2GenFun1}. (b) Deformation of $C$ gives the closed contours $C_1$, $C_2$, which enclose the $z_1$ and $z_2$ poles of $B(z;t)$, respectively, along with contour at infinity, $C_\infty$.}
\label{ContourFig}
\end{figure}

In this appendix we present the calculation of exponentially suppressed corrections in $p$, given in \eqref{N2LargeChargeCorrNP}, essentially following the method of \cite{Dorigoni:2022cua}. We start by writing \eqref{eq:spectral_decomp} as 
\be
\label{GmaxCorr2}
{\mathcal{G}}_p^{(N)}(\tau)=\frac12 \left\langle{\mathcal{G}}_p^{(N )}\right\rangle+\frac{1}{4 \pi i} \int_{\operatorname{Re} s=1+\eps} d s \frac{\pi}{\sin (\pi s)} s(1-s)(2 s-1)^2 {g}_p^{(N  )}(s) E_s^*(\tau)\,,
\ee
where we shifted the integration contour just past the pole at $s=1$. This is done for computational convenience. We begin by writing the Eisenstein series as a lattice integral 
\be
E_s^*(\tau) = \frac{1}{2} \sum_{(m,n)\neq (0,0)} \int_0^\infty dt e^{-t Y_{m n}(\tau)} t^{s-1}~.
\ee
Consider the following generating function
\be \label{GeneratingFuncP}
{\mathcal{G}}^{(N )}(z;\tau) := \sum_{p=2,4,6, \cdots}{\mathcal{G}}_p^{(N )}(\tau) z^{p/2}~,
\ee
Performing the sum over $p$, one finds
\begin{align}
\label{EqC9}
{\mathcal{G}}^{(N)}(z;\tau) &= -\frac{z N(N-1)}{4\left(N^4-1\right) \left(z-1\right)} \[N^2+1+\left(N^2-1\right) z \, _2F_1\left(1,\frac{N^2+1}{2};\frac{N^2+3}{2} ;z\right)\]\no\\[5pt]
&-\frac14 \sum_{(m,n)\neq (0,0)} \int_0^\infty dt e^{-t Y_{m n}(\tau)} \sum_{n=2}^\infty \underset{{s=n}}{\operatorname{Res}} \[\frac{\pi}{\sin (\pi s)} s(1-s)(2 s-1)^2 {g}^{(N )}(z;s)t^{s-1}\]~.
\end{align}
The first line comes from summing over the average term in \eqref{GmaxCorr2}. In the second line we did the contour integral by summing over residues. The term ${g}^{(N )}(z;s)$ is given by
\begin{align}\label{eqeqeq}
{g}^{(N )}(z;s)&:=\sum_{p=2,4,6,\cdots} {g}_p^{(N )}(s)z^{p/2} \\&= \frac{N(N-1)}{2s(1-s)} {}_3F_2(2-N,s,1-s;3,2;1) \frac{1}{z-1}\[\, _2F_1\left(1-s,s;\frac{1}{2} \left(N^2-1\right);\frac{z}{z-1}\right)-1\]
\end{align}
In computing the above sum, one uses the following summation identity
\be
\sum_{k=0}^{\infty} \frac{\left(a_1\right)_k}{k !}{ }_3 F_2\left(-k, a_2, a_3 ; b_1, b_2 ; w\right) z^k=\left(\frac{1}{1-z}\right)^{a_1}{ }_3 F_2\left(a_1, a_2, a_3 ; b_1, b_2 ; \frac{z w}{z-1}\right)~.
\ee
Computing the sum over residues in second line of \eqref{EqC9} is rather involved due to the product of hypergeometrics in \eqr{eqeqeq}. Therefore, for simplicity we take $N=2$ henceforth. In this case the generating function is
\begin{align}\label{N=2GenFun}
{\mathcal{G}}^{(2  )}(z;\tau) &=  -\frac{\sqrt{z}-\tanh ^{-1}(\sqrt{z})}{2\sqrt{z}(1-z)}-\frac14 \sum_{(m,n)\neq (0,0)} \int_0^\infty dt ~e^{-t Y_{m n}(\tau)} B(z;t)
\end{align}
where
\be
\label{BFunction}
B(z;t) =\frac{z \big(4 (t-3) t (t+1)^2 (3 t-1)-4 (t-1)^2 t (t (3 t-2)+3) z\big)}{(t+1)^3 \left(z-1\right) \big((t+1)^2-(t-1)^2 z\big)^2}~.
\ee
For future reference we note that this function has poles in $z$ at
\be
z_1= 1~,\qquad z_2= \frac{(t+1)^2}{(t-1)^2}~,
\ee
where $z_1$ is a simple pole and $z_2$ is a double pole. Notice that the two terms in \eqref{N=2GenFun} can be conveniently combined together:
\begin{align}
\label{randomEq}
{\mathcal{G}}^{(2)}(z;\tau) &=  -\frac14 \sum_{(m,n)\in \mathbb{Z}^2} \int_0^\infty dt ~e^{-t Y_{m n}(\tau)} B(z;t)
\end{align}
which follows from the fact that the $t$-integral over $B(z;t)$ (for $0<z<1$) is given by
\be
-\frac14 \int_0^\infty dt~ B(z;t)= - \frac{\sqrt{z}-\tanh ^{-1}\left(\sqrt{z}\right)}{2 \sqrt{z} (1-z)} =  \frac{1}{4} \sum_{p=2,4,6,\cdots} \left(H_{\frac{p+1}{2}}-H_{\frac{1}{2}}\right)z^{p/2}~.
\ee
The $B(z;t)$ function satisfies the following inversion property
\be
B(z;t) = t^{-1} B(z;t^{-1})~,
\ee
from which it follows that the $t$-integral in \eqref{randomEq} can be written as  \cite{Dorigoni:2022cua}
\begin{align}
{\mathcal{G}}^{(2)}(z;\tau) &= - \frac12 \sum_{(m,n)\in \mathbb{Z}^2} \int_1^\infty dt ~e^{-t Y_{m n}(\tau)} B(z;t)~.
\end{align}
Terms of fixed charge $p$ can be picked off by the following integral transform
\begin{align}
\label{N=2GenFun1}
{\mathcal{G}}_{p}^{(2 )}(\tau) &= - \frac12 \sum_{(m,n)\in \mathbb{Z}^2} \int_1^\infty dt ~e^{-t Y_{m n}(\tau)} \left[\oint_C \frac{B(z;t)}{z^{\frac p2+1}} \frac{d z}{2 \pi i}\right]~.
\end{align}
where $C$ is the closed contour enclosing the origin of the complex $z$ plane as shown in Figure \ref{ContourFig}.

We are interested in the large $p$ behaviour of the above expression. To extract this behaviour we first deform the integration contour $C$ past infinity and pick up the contributions from the poles at $z_1$ and $z_2$ as shown in Figure \ref{ContourFig} above. (This situation is somewhat simpler than \cite{Dorigoni:2022cua} where one encountered a branch cut between $z_1$ and $z_2$.) There is no contribution from the contour $C_\infty$ at infinity for any $p>0$ because $B(z;t)\sim 1/z$ as $z\to\infty$. 

The remaining contours $C_1$ and $C_2$ contribute as follows:
\begin{align}
\label{N=2Fun1}
{\mathcal{G}}_{p}^{(2)}(\tau) &=  -\frac12 \sum_{(m,n)\in \mathbb{Z}^2} \int_1^\infty dt ~e^{-t Y_{m n}(\tau)} \left[-\frac{t^2-6 t+1}{(t+1)^3}\right]\no\\[5pt]
&-\frac12 \sum_{(m,n)\in \mathbb{Z}^2} \int_1^\infty dt ~e^{-t Y_{m n}(\tau)} \left[\frac{t^2-2 (2p +3) t +1 }{(t+1)^3}\right]z_2^{-\frac p2}
\end{align}
The $p$-dependence lies in the second line. Recalling that $z_2$ depends on $t$, one finds that the $t$ integral in the second line is dominated by the following saddle points
\be
t_1^\star = \sqrt{\frac{{Y_{m n}(\tau)+ 2p }}{{Y_{m n}(\tau)}}}=- t_2^\star ~.
\ee
Since the domain of the $t$-integral is $[1,\infty)$, we need only consider the $t_1^\star$ saddle. Consequently, the saddle-point evaluation of the exponential term in the second line of \eqref{N=2Fun1} is
\be
\exp \left[-t_1^{\star} Y_{m n}(\tau)- p \log \left(\frac{t_1^{\star}+1}{t_1^{\star}-1}\right)\right]~\stackrel{p \rightarrow \infty}{\sim}~ \exp \left(-2 \sqrt{2p Y_{mn}(\tau)}\right)
\ee
This determines the parametric scale of leading exponentially small corrections in $p$. To determine the power law coefficient we integrate over small fluctuations around the saddle point $t_1^\star$. To this end, define
\be
S(t):=t Y_{m n}(\tau)+ p \log \left(\frac{t+1}{t-1}\right)
\ee
and denote the fluctuation of $t$ around $t_1^\star$ by $t = t_1^\star + p^{\frac14} \delta $. The quadratic fluctuation of $S(t)$ is given by
\be
S\left(t_1^{\star}+p^{\frac{1}{4}} \delta\right)=S\left(t_1^{\star}\right)+k^2 \delta^2+O\left(p^{-\frac{1}{4}} \delta^3\right)~,\qquad k^2:=\frac{Y_{m n}^{3/2}(\tau)}{\sqrt{2}}\left(1+O\left(p^{-1}\right)\right)~.
\ee
Plugging this expression into \eqref{N=2Fun1} and performing the Gaussian integral over $\delta$ we get the following result for the leading exponentially suppressed term:
\begin{align}
\sum_{(m,n)\in \mathbb{Z}^2} \exp \left(-2 \sqrt{2p Y_{m n}(\tau) }\right)\[\(2\pi^2pY_{m n}(\tau)\)^{1/4}+O(p^{-1/4})\] 
\end{align}
Noting that $Y_{00}(\t)=0$, and using the definition \eqref{DFunction} for the $SL(2,\mathbb{Z})$-invariant function $D_{p/2}(r;\tau)$, our final answer for the leading non-perturbative corrections to $\cG_p^{(2)}(\tau)$ in the large charge expansion takes the form \eqref{N2LargeChargeCorrNP}.

\section{The spectral representation of $D_N(r;\tau)$}\label{app:D_spectral_rep}

Here we derive the $SL(2,\Z)$ spectral representation of the function $D_N(r;\t)$.\foot{This calculation was independently performed by Scott Collier, whom we thank for a discussion.} This non-holomorphic function, which is square-integrable and $SL(2,\Z)$-invariant, was introduced in \cite{Dorigoni:2022cua}. The function $D_{p/2}(r;\t)$ appeared in \eqr{N2LargeChargeCorrNP} and \eqr{N3LargeChargeCorrNP} as the non-perturbative function in the large $p$, finite $N$ limit of $\cG_p^{(N)}(\t)$.

A definition of this function was given in \eqr{DFunction}. An alternative definition, more useful for present purposes, is as a $PSL(2,\Z)$ Poincar\'e sum over images \cite{Dorigoni:2022cua}:
\e{}{D_N(r;\t) = 2 \sum_{\g\in PSL(2,\Z)/\Gamma_\i} d_N(r;\g\t)\,,\quad d_N(r;y) := \pi^{-r}y^r\, \Li_{2r}\(e^{-4\sqrt{\pi N/ y}}\)}
The ``seed'' function $d_N(r;\t)$ being $T$-invariant, we have modded out by $\G_\i$, the set of modular $T^n$ transformations.

We want to compute the (bracketed) overlap with the Eisenstein series, 
\e{}{\{D_N(r),E_{1-s}\} = \L(1-s)^{-1}(D_N(r),E_{1-s}) }
The inner product $(D_N(r),E_{1-s})$ is defined by the Rankin-Selberg transform of $D_N(r;\t)$. The computation turns out to be simple because we can use the ``unfolding trick'' on $D_N(r;\t)$, leading to a straightforward one-dimensional integral against $E_{s,0}(y)$, the zero mode of the Eisenstein series. Before proceeding, we note that, by the same logic, the overlap of $D_N(r;\t)$ with the Maass cusp forms vanishes, because the latter have no zero mode:
\e{}{(D_N(r),\phi_n)=0~~\forall~n.}

Proceeding with the Eisenstein overlap, we have
\e{}{\{D_N(r),E_{1-s}\} = {2\o \L(1-s)} \int_0^\i {dy\o y^2} \,d_N(r;y) \,y^{s} + (s \rar 1-s)}
Now use
\e{}{\Li_{2r}(x) = \sum_{k=1}^\i {x^{k}\o k^{2r}}}
Swapping the sum and integral and performing them in succession is elementary. The integral yields\foot{Convergence of the integral near the cusp requires $\Re(r+s)<1$.}
\e{}{\{D_N(r),E_{1-s}\} = {\pi ^{s-1} 4^{2r+2s-1} N^{r+s-1} \Gamma (2 (1-r-s))\o  \L(1-s)} \sum_{k=1}^\i  k^{2(s-1)} + (s \rar 1-s)}
Performing the sum leads to the final result,
\e{Doverlap}{{\{D_N(r),E_{1-s}\} = { 4^{1+2r-2s} N^{r-s} \Gamma (2 (s-r))\o \G(s)} + (s \rar 1-s)}}
Note that while $D_N(r;\t)$ parameterizes exponentially suppressed terms at $p\gg1$, the spectral overlap itself is not exponential {\it per se}. Rather, the sign of such terms is the fact the overlap is factorially divergent at $s\rar\pm\i$.

For generic $r$, the modular average of $D_N(r;\t)$ is
\es{}{\< D_N(r) \> &:= \int_\cF {dx dy \o y^2} D_N(r;\t)\\ &= \half \lim_{s\rar 1} \{D_N(r),E_{1-s}\}\\
&= 2^{4r-3} N^{r-1} \Gamma (2 -2r) \qquad (2r \notin \Z_{\geq 0})}
For $2r\in \Z_{\geq 0}$, the result is finite but takes a different form which may be easily extracted from \eqr{Doverlap}. The full spectral decomposition is therefore
\e{}{D_N(r;\t) = \< D_N(r) \> + {1\o 2\pi i}\int_{\Re s = \half} ds \,{ 4^{1+2r-2s} N^{r-s} \Gamma (2 (s-r))\o \G(s)} E_{s}^*(\t)}
\subsubsection*{'t Hooft limit}

Using the above result, we now show that the non-perturbative corrections in the strong 't Hooft coupling expansion \eqref{N2thooftNP} can be derived directly from the non-perturbative corrections in the large charge expansion at finite coupling given in \eqref{N2LargeChargeCorrNP}. We begin by writing the leading term in $\cF^{(2)}_{\text {NP}}(p; \tau)$ in the spectral form
\begin{align}
\label{D30}
\big(2\pi^2 p\big)^{1 / 4} D_{p/ 2}\left(-\frac{1}{4}; \tau\right) =&~ \frac{3 \pi }{16 \sqrt{2} p} + {1\o 2\pi i}\int_{\Re s = \half} ds \frac{\sqrt{\pi } 2^{\frac{3}{2}-3 s} p^{-s} \Gamma \left(2 s+\frac{1}{2}\right)}{\Gamma (s)} \(\Lambda(s)y^s+\Lambda(1-s)y^{1-s}\)\no\\
&+\text{(non-zero modes)}\,.
\end{align}
Shifting the contour slightly to $\Re s=\half-\eps$ for an infinitesimal $\eps>0$, taking the double-scaling limit of fixed $\l_p$, and keeping only the leading order term $\sim p^0$, we get 
\begin{align}
\big(2\pi^2 p\big)^{1 / 4} D_{p/ 2}\left(-\frac{1}{4}; \tau\right) \approx  {1\o 2\pi i}\int_{\Re s = \half-\eps} ds \sqrt{\pi } 2^{\frac{3}{2}-s} \lambda_p^{-s} \zeta (2 s) \Gamma \left(2 s+\frac{1}{2}\right),
\end{align}
which comes from the first term in the parenthesis of \eqref{D30}. The integrand has poles at $s= -\frac14(2n+1)$ for all $n\in\Z_{\geq 0}$, and decays exponentially at infinity for $\l_p < 2\pi^2$. Deforming the contour to the left and using the series representation of $\zeta(2s)$ we find that the sum over the residues is given by
\begin{align}
{1\o 2\pi i}\int_{\Re s = \half-\eps} ds \sqrt{\pi } 2^{\frac{3}{2}-s} \lambda_p^{-s} \zeta (2 s) \Gamma \left(2 s+\frac{1}{2}\right) &= \sqrt{2\pi}(2\lambda_p)^{1/4} \sum_{k=1}^\infty \sqrt{k}  e^{-\sqrt{2} k \sqrt{\lambda_p}}\no\\
&=\sqrt{2\pi}(2\lambda_p)^{1/4}  \text{Li}_{-\frac{1}{2}}\left(e^{-\sqrt{2\lambda_p} }\right).
\end{align}
This matches the result obtained in \eqref{N2thooftNP} for the leading-order non-perturbative corrections in $\l_p\gg1$.

\section{Resurgence of the $\l_p\gg1$ expansion at $N=3$}\label{app:resurgence}
In this appendix we apply resurgence to the strong coupling expansion \eqref{eq:strong_coupling_expansion_N3} to compute the exponentially small non-perturbative terms quoted in \eqref{n3resurgence}. Following \cite{Dorigoni:2021guq}, we consider the modified Borel transform of the infinite sum in \eqref{eq:strong_coupling_expansion_N3},\footnote{For convenience, here we use the Borel transform with rescaled argument $\sqrt{2}w$, compared to $2w$ as in \cite{Dorigoni:2021guq}.}
\begin{align}\label{eq:Borel_resummation}
	\mathcal{B}: \quad\sum_{n=1}^\infty c_3(n)\,x_p^{-2n-1} \quad\mapsto\quad \phi(w):=\sum_{n=1}^\infty \frac{2\pi c_3(n)}{\zeta(2n+1)\Gamma(2n+2)}(\sqrt{2}w)^{2n+1}\,,
\end{align}
where we have introduced $x_p=\sqrt{\lambda_p}$ and the coefficients $c_3(n)$ have been given previously in \eqref{eq:coeffs_c3}. For the Borel transform \eqref{eq:Borel_resummation} we then find
\begin{align}
\begin{split}\label{eq:phi_w}
	\phi(w) = \frac{9}{35} \pi  w^3 \Big[&280\,_3F_2\big(-\tfrac{3}{2},\tfrac{3}{2},\tfrac{3}{2}\big|1,\tfrac{5}{2}\big|w^2\big)-168 w^2 \, _3F_2\big(-\tfrac{1}{2},\tfrac{5}{2},\tfrac{5}{2}\big|2,\tfrac{7}{2}\big|w^2\big)\\
	&+25 w^4 \, _3F_2\big(\tfrac{1}{2},\tfrac{7}{2},\tfrac{7}{2}\big|3,\tfrac{9}{2}\big|w^2\big)\Big]\,,
\end{split}
\end{align}
which is input to the directional Borel resummation of \eqref{eq:strong_coupling_expansion_N3} given by
\begin{align}
	\mathcal{S}_\theta{\mathcal{G}}_{\mathfrak{g}=0}^{(3)}(x_p) = \frac{3}{2}\log\Big(\frac{x_p^2}{32\pi^2}\Big)+\frac{7}{2}+3\gamma_{E} + \frac{x_p}{\sqrt{2}\pi}\int_0^{e^{i\theta}\infty}\frac{dw}{4\sinh^2(x_p w/\sqrt{2})}\,\phi(w)\,,
\end{align}
for some $\theta\in(-\frac{\pi}{2},\frac{\pi}{2})$. However, due to the branch cut along $[1,\infty)$ of the hypergeometric functions in $\phi(w)$, the result of the Borel resummation is ambiguous, and in fact not real for $x_p>0$. This suggests that the asymptotic expansion is not Borel summable and hence requires exponentially small corrections. Such terms are encoded in the so-called `Stokes automorphism', defined as the difference between the two lateral resummations just above and below the branch cut of $\phi(w)$,
\begin{align}\label{eq:stokes_automorphism}
	\delta\mathcal{G}_{\mathfrak{g}=0}^{(3)}(x_p) := (\mathcal{S}_+-\mathcal{S}_-)\mathcal{G}_{\mathfrak{g}=0}^{(3)}(x_p) = \frac{x_p}{\sqrt{2}\pi}\int_1^\infty \frac{dw}{4\sinh^2(x_p w/\sqrt{2})}\,\text{disc}\big[\phi(w)\big],
\end{align}
where $\mathcal{S}_{\pm}=\lim_{\theta\to\theta^{\pm}}\mathcal{S}_\theta\mathcal{G}_{\mathfrak{g}=0}^{(3)}(x_p)$ and we already anticipated that the $w$-integration actually starts from 1 (since $\text{disc}\big[\phi(w)\big]=0$ for $0\leq w\leq 1$).

To compute the discontinuity of $\phi(w)$ across the branch cut,
\begin{align}
	\text{disc}\big[\phi(w)\big] := \phi(w+i0) - \phi(w-i0)\,,
\end{align}
we make use of Euler's integral transform for the hypergeometric $_3F_2$'s appearing in \eqref{eq:phi_w}:
\begin{align}\label{eq:eulers_integral_trafo}
	_3F_2\big(a_1,a_2,c\big|b_1,d\big|z\big) = \frac{\Gamma(d)}{\Gamma(c)\Gamma(d-c)}\int_{0}^{1}dt~t^{c-1}(1-t)^{d-c-1}\,_2F_1\big(a_1,a_2;b_1;tz\big)\,.
\end{align}
The only piece of the integrand which has a branch cut is the $_2F_1$, whose discontinuity is given by the formula
\begin{align}\label{eq:disc_2F1}
\begin{split}
	\text{disc}\big[{_2F_1}\big(a,b;c;z\big)\big] = \frac{2\pi i\,\Gamma(c)}{\Gamma(a)\Gamma(b)\Gamma(c-a-b+1)} &z^{1-c}(z-1)^{c-a-b}\\
	&\times{_2F_1}\big(1-a,1-b;c-a-b+1;1-z\big)\,.
\end{split}
\end{align}
Note that since the discontinuity of the hypergeometric function starts at argument 1, when plugging \eqref{eq:disc_2F1} into \eqref{eq:eulers_integral_trafo} the range of the $t$-integration is modified to $1/z\leq t\leq 1$.\footnote{
	In the computation of the discontinuity of the last $_3F_2$ in \eqref{eq:phi_w}, after using \eqref{eq:eulers_integral_trafo}, one finds that formula \eqref{eq:disc_2F1} diverges. To avoid this divergence, we first use the shift identity
	\begin{align}
		{_3F_2}\big(\tfrac{1}{2},\tfrac{7}{2},\tfrac{7}{2}\big|3,\tfrac{9}{2}\big|w^2\big)=8\, {_3F_2}\big(-\tfrac{1}{2},\tfrac{7}{2},\tfrac{7}{2}|3,\tfrac{9}{2}|w^2\big)-7 \,{_2F_1}\big(-\tfrac{1}{2},\tfrac{7}{2};3;w^2\big)\,,
	\end{align}
	which then leads to a manifestly finite answer in the computation of the discontinuity.
}

We find that the discontinuity of $\phi(w)$ can be expressed as
\begin{align}\label{eq:disc_phi_w}
\begin{split}
	\text{disc}\big[\phi(w)\big]&=\frac{48 i}{\pi} \Bigg[-3 w G_{2,2}^{2,2}\left(w^2\left|
	\begin{array}{c}
	 \frac{1}{2},\frac{5}{2} \\
	 1,1 \\
	\end{array}
	\right.\right)-3 w G_{3,3}^{2,3}\left(w^2\left|
	\begin{array}{c}
	 \frac{1}{2},\frac{1}{2},\frac{7}{2} \\
	 1,2,-\frac{1}{2} \\
	\end{array}
	\right.\right)\\
	&\qquad\quad~\,+3 w G_{3,3}^{2,3}\left(w^2\left|
	\begin{array}{c}
	 \frac{1}{2},\frac{1}{2},\frac{7}{2} \\
	 2,2,-\frac{1}{2} \\
	\end{array}
	\right.\right)-4 w G_{3,3}^{2,3}\left(w^2\left|
	\begin{array}{c}
	 \frac{1}{2},\frac{1}{2},\frac{9}{2} \\
	 1,3,-\frac{1}{2} \\
	\end{array}
	\right.\right)\\
	&\qquad\quad~\,+3 G_{2,2}^{2,2}\left(1\left|
	\begin{array}{c}
	 -\frac{1}{2},\frac{3}{2} \\
	 0,0 \\
	\end{array}
	\right.\right)+3 G_{3,3}^{2,3}\left(1\left|
	\begin{array}{c}
	 -\frac{1}{2},-\frac{1}{2},\frac{5}{2} \\
	 0,1,-\frac{3}{2} \\
	\end{array}
	\right.\right)\\
	&\qquad\quad~\,+4 G_{3,3}^{2,3}\left(1\left|
	\begin{array}{c}
	 -\frac{1}{2},-\frac{1}{2},\frac{7}{2} \\
	 0,2,-\frac{3}{2} \\
	\end{array}
	\right.\right)-3 G_{3,3}^{2,3}\left(1\left|
	\begin{array}{c}
	 -\frac{3}{2},-\frac{3}{2},\frac{3}{2} \\
	 0,0,-\frac{5}{2} \\
	\end{array}
	\right.\right)\\
	&\qquad\quad~\,+\pi ^2 w^3 \, _2F_1\left(-\frac{5}{2},\frac{3}{2};1;1-w^2\right)\Bigg],
\end{split}
\end{align}
where $w>1$ and $G_{p,q}^{m,n}\left(z\left|\begin{array}{c}\{a_i\}\\\{b_i\}\end{array}\right.\right)$ denotes the Meijer-G function.

This is then plugged into \eqref{eq:stokes_automorphism}, which can be computed by first shifting the integration variable $w\mapsto w+1$ and expanding the $\sinh^2(x_pw/\sqrt{2})$ in the denominator:
\begin{align}
	\delta\mathcal{G}_{\mathfrak{g}=0}^{(3)}(x_p) = \frac{\tilde{x}_p}{\pi}\sum_{n=1}^\infty n\,e^{-2n\tilde{x}_p}\int_0^\infty d{w} ~ e^{-2n\tilde{x}_p{w}}\,\text{disc}\big[\phi({w}+1)\big]\,,
\end{align}
where for convenience we introduced $\tilde{x}_p=x_p/\sqrt{2}=\sqrt{\lambda_p/2}$. The integration over ${w}$ can be done by expanding the discontinuity around $w=0$ and using the integral identity $\int_0^\infty dw~e^{-2n\tilde{x}_pw}w^m=\frac{m!}{(2n\tilde{x}_p)^{m+1}}$ term by term. The series expansion of the discontinuity \eqref{eq:disc_phi_w} is of the form
\begin{align}
	\text{disc}\big[\phi(w+1)\big]=\pi i\,\sum_{m=0}^\infty a_m w^m\,,
\end{align}
where the coefficients $a_m$ turn out to be rational numbers. The first few are given by
\begin{align}
	a_m=\Big\{48,\,408,\,1839,\,\frac{6737}{2},\,\frac{192939}{64},\,\frac{834603}{640},\,\frac{222319}{1024},\,\ldots\Big\},\qquad\text{for }m=0,1,2,\ldots\,.
\end{align}
Putting all together, we have
\begin{align}
\begin{split}
	\delta\mathcal{G}_{\mathfrak{g}=0}^{(3)}(x_p) &= i\tilde{x}_p \sum_{n=1}^\infty n\,e^{-2n\tilde{x}_p}\Big[\frac{48}{(2n\tilde{x}_p)}+\frac{408}{(2n\tilde{x}_p)^2}+\frac{3678}{(2n\tilde{x}_p)^3}+\frac{20211}{(2n\tilde{x}_p)^4}+\frac{578817}{8(2n\tilde{x}_p)^5}+\ldots\Big]\\
	&= i\sum_{n=1}^\infty e^{-2n\tilde{x}_p}\Big[24+\frac{102}{n\tilde{x}_p}+\frac{1839}{4(n\tilde{x}_p)^2}+\frac{20211}{16(n\tilde{x}_p)^3}+\frac{578817}{256(n\tilde{x}_p)^4}+\ldots\Big],
\end{split}
\end{align}
where each term in this sum corresponds to a perturbative contribution in $1/\tilde{x}_p\sim1/\sqrt{\lambda_p}$ to a non-perturbative correction of order $e^{-2n\tilde{x}_p}=e^{-n\sqrt{2\lambda_p}}$. Note that this is in exact agreement with the scale $\Lambda^2(\lambda_p)$ predicted by the weak coupling radius of convergence, c.f. \eqref{npscale}.

Term by term in $1/\tilde{x}_p$, the sum over $n$ resums into polylogarithms and we have
\begin{align}\label{n3resurgence_2}
\begin{split}
	\delta\mathcal{G}_{\mathfrak{g}=0}^{(3)}(\lambda_p) = i\,\bigg[&24\,\text{Li}_0\big(e^{-\sqrt{2\lambda_p}}\big)+\frac{102\sqrt{2}}{\lambda_p^{1/2}}\,\text{Li}_1\big(e^{-\sqrt{2\lambda_p}}\big)+\frac{1839}{2\lambda_p}\,\text{Li}_2\big(e^{-\sqrt{2\lambda_p}}\big)\\
	&+\frac{20211}{4\sqrt{2}\lambda_p^{3/2}}\,\text{Li}_3\big(e^{-\sqrt{2\lambda_p}}\big)+\frac{578817}{64\lambda_p^2}\,\text{Li}_4\big(e^{-\sqrt{2\lambda_p}}\big)+\ldots\bigg],
\end{split}
\end{align}
which is the result already quoted in the main text.

We can now assemble the median Borel resummation, $\mathcal{S}_{\text{med}}:=\frac{1}{2}\big(\mathcal{S}_+-\mathcal{S}_-\big)$, by adding the discontinuity \eqref{n3resurgence_2} to one of the lateral resummations:
\begin{align}\label{median}
\begin{split}
	\mathcal{S}_{\text{med}}\mathcal{G}_{\mathfrak{g}=0}^{(3)}(\lambda_p) &= \mathcal{S}_{\pm}\mathcal{G}_{\mathfrak{g}=0}^{(3)}(\lambda_p)\mp\frac{1}{2}\delta \mathcal{G}_{\mathfrak{g}=0}^{(3)}(\lambda_p)\\
	&= \frac{3}{2}\log\Big(\frac{\lambda_p}{32\pi^2}\Big)+\frac{7}{2}+3\gamma_{E}+\frac{\sqrt{\lambda_p}}{\sqrt{2}\pi}\int_0^{\infty}\frac{dw}{4\sinh^2(w\sqrt{\lambda_p}/\sqrt{2})}~\text{Re}\big[\phi(w)\big]\,,
\end{split}
\end{align}
with $\phi(w)$ given in \eqref{eq:phi_w}. This is now a manifestly real expression for any $\lambda_p>0$, where the non-perturbative corrections are captured by the discontinuity $\delta\mathcal{G}_{\mathfrak{g}=0}^{(3)}(\lambda_p)$ given in \eqref{n3resurgence_2}.

\section{The genus-one correlator $\mathcal{G}_{\mathfrak{g}=1}^{(N)}(\lambda_p)$}\label{app:large_p_g1}

For the first sub-leading contribution at order $p^{-1}$ one finds
\begin{align}\label{eq:large_p_g1_v2}
\begin{split}
	\mathcal{G}_{\mathfrak{g}=1}^{(N)}(\lambda_p) &=-\frac{3N(N^2-1)\zeta(3)}{16\pi^2}\,\lambda_p +\frac{N(N^2-1)(N-1)}{8}\\[5pt]
	&\quad + {1\over 2\pi i} \int_{\Re s =\frac{1}{2}} ds \,{\pi\over \sin(\pi s)} s(1-s)(2s-1)^2\, \Lambda(1-s)\(\frac{\lambda_p}{4\pi}\)^{s-1} g_2^{(N)}(s)\, F^{(1)}(N,s)\,.
\end{split}
\end{align}
Its large $\lambda_p$ expansions reads
\begin{align}
\begin{split}
	\frac{\text{(2nd line of \eqref{eq:large_p_g1})}}{N^2-1} = \frac{\sqrt{\frac{\pi }{2}}\,\Gamma\big(\frac{N^2+1}{2}\big)}{\Gamma\big(\frac{N^2-4}{2}\big)}
	\bigg[-\frac{3\zeta(3)}{4\lambda_p^{\frac{3}{2}}}\,\widehat f_2^{(N)}(-\tfrac{1}{2})+\frac{45(N^2-6) \zeta(5)}{32\lambda_p^{\frac{5}{2}}}\,\widehat f_2^{(N)}(-\tfrac{3}{2})++O(\lambda_p^{-\frac{7}{2}})\bigg],
\end{split}
\end{align}
where the same comments about the truncation for $N$ even apply again. For the perturbative expansion at small $\lambda_p$ we have
\begin{align}
	\frac{{\mathcal{G}}_{\mathfrak{g}=1}^{(N)}(\lambda_p)}{N^2-1} = -\frac{75N^2\zeta(5)}{128\pi^4(N^2+1)}\,\lambda_p^2 + \frac{3675N^3\zeta(7)}{2048\pi^6(N^2+1)(N^2+3)}\,\lambda_p^3+O(\lambda_p^4)\,,
\end{align}
where the constant and linear terms in $\lambda$ have cancelled (i.e. the entire first line of \eqref{eq:large_p_g1} gets cancelled by the $s=1$ and $s=2$ residues from the second line).

\paragraph{Bessel integral representations:} Proceeding as in the $\mathfrak{g}=0$ case, for $N=2$ we find the expression
\begin{align}
	{\mathcal{G}}_{\mathfrak{g}=1}^{(2)}(\lambda_p)=-\int_0^{\infty}dw~\frac{3w^2}{4\pi^2\sinh^2(w)}~\bigg(w\lambda_p-\sqrt{2\lambda_p}\pi J_1\Big(\tfrac{w\sqrt{2\lambda_p}}{\pi}\Big)\bigg)\,.
\end{align}
The $N=3$ result takes again a more complicated form given by
\begin{align}
\begin{split}
	{\mathcal{G}}_{\mathfrak{g}=1}^{(3)}(\lambda_p)&=-\int_0^{\infty}dw~\frac{3}{w^3\lambda_p^2\pi^2\sinh^2(w)}~\bigg(w^6\lambda_p^3+16\pi^2w^2\lambda_p(24\pi^2-w^2\lambda_p)J_0^2\Big(\tfrac{w\sqrt{\lambda_p}}{\sqrt{2}\pi}\Big)\\
	&\qquad\qquad\qquad-4\sqrt{2\lambda_p}w\pi(384\pi^4-40\pi^2w^2\lambda_p+w^4\lambda_p^2)J_0\Big(\tfrac{w\sqrt{\lambda_p}}{\sqrt{2}\pi}\Big)J_1\Big(\tfrac{w\sqrt{\lambda_p}}{\sqrt{2}\pi}\Big)\\
	&\qquad\qquad\qquad+4(768\pi^6-128\pi^4w^2\lambda_p+5\pi^2w^4\lambda_p^2)J_1^2\Big(\tfrac{w\sqrt{\lambda_p}}{\sqrt{2}\pi}\Big)\bigg)\,.
\end{split}
\end{align}

\section{Derivation of Bessel integral representation in $p = \a N^2$ regime}\label{besselapp}
Here we resum the weak coupling expansion of $\mathcal{G}_\alpha^{(0)}(\lambda)$, the planar integrated correlator in the gravity regime $p = \a N^2 \gg 1$ with fixed $\a$, to derive the Bessel integral representation \eqr{besselrepfinal}. 

Using the zeta function integral identity \eqref{eq:zeta_integral_identity}, we can resum the weak coupling expansion \eqref{eq:large_p=N^2_weak}. As before, we split the computation again into two parts. For the first sum over $\widetilde h_0(s)$, we find 
\begin{align}\label{eq:part1}
	\widetilde h_0(s):\quad \int_0^\infty dw~ \frac{w}{\sinh^2(w)}\bigg(-\frac{1}{2}+\frac{2\pi^2}{\lambda w^2}J_1^2\Big(\tfrac{w\sqrt{\lambda}}{\pi}\Big)\bigg)\,.
\end{align}
To perform the second sum over $\widetilde h_\alpha(s)$ we make use of the integral representation $P_n(z)=\frac{1}{\pi}\int_0^{\pi}dt\,(z-\sqrt{z^2-1} \cos (t))^n$ for the Legendre polynomial. This yields
\begin{align}
	\widetilde h_\alpha(s):\qquad \int_0^\infty dw\int_0^\pi dt~\frac{w}{\sinh^2(w)}\Bigg(\frac{1}{2\pi}-\frac{2\pi}{\lambda w^2}\frac{J_1\({w \sqrt{\lambda u_\a}\o \pi}\)^2}{u_\a}\Bigg)\,,
\end{align}
where $u_\a:=1+2\a-2\sqrt{\alpha(\alpha+1)}\cos(t)>0$ for $\a\geq 0$. Performing the $t$-integral for the first term in the bracket gives $1/2$, which then cancels against the first term from \eqref{eq:part1} when adding the two parts. Altogether, 
\begin{align}\label{g0eq}
	\mathcal{G}_\alpha^{(0)}(\lambda) = \frac{2\pi^2}{\lambda}\int_0^\infty dw~{J_1^2\({w \sqrt{\lambda}\o \pi}\)-K_{11}\({w \sqrt{\lambda}\o\pi};\a\)\o w \sinh^2(w)}\,,
\end{align}
where we have defined the localized\foot{More general Bessel kernels may be defined as functions of two positions $(x,x')$.} integral Bessel kernel $K_{11}(x;\a)$ in \eqr{Kdef}. Note that $dt = {du_\a\o \sqrt{u_\a(4\a+2-u_\a)-1}}$. Since $u_0=1$, the first term in \eqr{g0eq} is simply the second term at $\a=0$, which allows us to nicely combine them:
\begin{align}
	\mathcal{G}_\alpha^{(0)}(\lambda) = {2\pi^2\o\lambda}\int_0^\infty dw~{K_{11}\({w \sqrt{\lambda}\o\pi};0\) - K_{11}\({w \sqrt{\lambda}\o\pi};\a\)\o w \sinh^2(w)}\,.
\end{align}
This is the final result \eqr{besselrepfinal}. 

We point out that this vanishes linearly as $\a\rar0$, as is manifest from the spectral representation \eqr{eq:h0}. Near $\a=0$, 
\es{smallalphaexpansion}{\mathcal{G}_\alpha^{(0)}(\lambda) &= \a\,\int_0^\infty dw \,w \,{J_1\big(\frac{w \sqrt{\lambda }}{\pi }\big)^2-J_2\big(\frac{w \sqrt{\lambda }}{\pi }\big)^2\o \sinh^2(w)}\\&~-\, \a^2\int_0^\infty dw\,\frac{\l w^3}{4\pi^2}\({J_1\big(\frac{w \sqrt{\lambda }}{\pi }\big)^2-J_2\big(\frac{w \sqrt{\lambda }}{\pi }\big)^2+ {\pi\o\sqrt{\lambda} w} J_1\big(\frac{w \sqrt{\lambda }}{\pi }\big) J_2\big(\frac{w \sqrt{\lambda }}{\pi }\big)\o \sinh^2(w)}\)+ O(\a^3)\,.}
The first term is precisely the $p=2$ integrated correlator in the planar 't Hooft limit \cite{Collier:2022emf}; this correspondence was anticipated from \eqr{eq:h0} expanded near $\a = 0$, 
\e{}{h^{(0)}_\a(s) = {\a\o2}\,g^{(0)}_2(s) + O(\a^2)\,.}
At $O(\a^2)$ and beyond, one finds a three-term linear combination of the same general form as the $O(\a^2)$ result, with the same Bessel functions appearing, dressed by increasing powers of $w \sqrt{\lambda}/\pi$.

\bibliographystyle{JHEP}
\bibliography{large_p}
\end{document}